\documentclass[sigconf]{acmart}

\AtBeginDocument{%
  \providecommand\BibTeX{{%
    \normalfont B\kern-0.5em{\scshape i\kern-0.25em b}\kern-0.8em\TeX}}}

\setcopyright{acmcopyright}
\copyrightyear{2020}
\acmYear{2020}
\acmDOI{10.1145/1122445.1122456}

\acmConference[ICONS '20]{International Conference on Neuromorphic Systems '20: }{July 28--30, 2020}{Virtual Conference}
\acmPrice{15.00}
\acmISBN{978-1-4503-XXXX-X/18/06}

\begin{document}

\title{Sequence learning in Associative Neuronal-Astrocytic Networks}

\author{Leo Kozachkov}
\affiliation{%
  \institution{Computational Brain Lab}
  \city{Rutgers University}
   \state{New Jersey, USA}
     \authornote{Leo Kozachkov is now with the Department of Brain and Cognitive Sciences, Massachusetts Institute of Technology, Cambridge, MA, USA.}
}


\author{Konstantinos P. Michmizos}
\affiliation{%
  \institution{Computational Brain Lab}
  \city{Rutgers University}
     \state{New Jersey, USA}
}
\email{konstantinos.michmizos@cs.rutgers.edu}

\renewcommand{\shortauthors}{Kozachkov \& Michmizos}

\begin{abstract}
The neuronal paradigm of studying the brain has left us with limitations in both our understanding of how neurons process information to achieve biological intelligence and how such knowledge may be translated into artificial intelligence and its most brain-derived branch, neuromorphic computing. Overturning our fundamental assumptions of how the brain works, the recent exploration of astrocytes is revealing that these long-neglected brain cells dynamically regulate learning by interacting with neuronal activity at the synaptic level. Following recent experimental evidence, we designed an associative, Hopfield-type, neuronal-astrocytic network and analyzed the dynamics of the interaction between neurons and astrocytes. We show that astrocytes were sufficient to trigger transitions between learned memories in the neuronal component of the network. Further, we mathematically derived the timing of the transitions that was governed by the dynamics of the calcium-dependent slow-currents in the astrocytic processes. Overall, we provide a brain-morphic mechanism for sequence learning that is inspired by, and aligns with, recent experimental findings. To evaluate our model, we emulated astrocytic atrophy and showed that memory recall becomes significantly impaired after a critical point of affected astrocytes was reached. This brain-inspired and brain-validated approach supports our ongoing efforts to incorporate non-neuronal computing elements in neuromorphic information processing.

\end{abstract}

\ccsdesc[500]{Computing methodologies~Self-organization}

\keywords{Associative Networks, Astrocytes, Sequence Learning}


\maketitle

\section{Introduction}
Understanding intelligence is a fundamental goal in several disciplines. Translating the understanding of biological intelligence to machines is a fundamental problem in Computing \cite{RN_3, RN_1, RN_2}. The breadth of solutions now offered by deep learning \cite{RN_4} has established the connectionist modeling of neural computation \cite{RN_8, RN_7, RN_5, RN_6} as the most faithful representation of the brain’s intelligence. Yet, despite their impressive performance, artificial neural nets are challenged by their intrinsic limitations in real-world applications \cite{RN_9}. Neural nets have a long way to go until they become computationally and energy efficient  \cite{RN_10}, tolerate variability in their input \cite{RN_11, RN_13, RN_12}, or separate learning from inference \cite{RN_15, RN_14, RN_13} –- tasks that brain networks are well-suited to execute by being radically different from the deep learning networks \cite{RN_16}. 

Neural connectionist algorithms are better fit for large-scale neuromorphic chips \cite{RN_22, RN_21, RN_17, RN_18, RN_19, RN_20} that are designed to run a time-dependent computational formalism, spiking neural networks (SNN) \cite{RN_23}, where asynchronous computing units are emulated as spiking neurons \cite{RN_24, RN_25, RN_26} and memory is distributed in the synapses. Indeed, by following a more faithful representation of the brain’s computational principles, we and others have used this non-Von Neumann architecture to reinforce the SNN’s promising results \cite{RN_33, RN_27, RN_29, RN_32, RN_31, RN_28, RN_30} with robustness \cite{RN_34}, and control our robots \cite{RN_35, RN_36, RN_39, RN_38, RN_40, RN_37, RN_41, RN_42} with energy-efficiency \cite{RN_42} and higher than the state-of-the-art accuracy \cite{tang2020reinforcement}. The main criticism to neuromorphic solutions is that, in the absence of fundamental algorithmic contributions, these promising results do not currently share the same scaling abilities with the mainstream deep learning approaches. To address this point, one alternative is to further pursue their biological plausibility by introducing new brain principles currently under study at the forefront of neuroscience\cite{RN_43}. Failure to do so means that biological principles now known to be critical for intelligence will remain out of reach for neuromorphic frameworks.

Since Ramon Y Cajal’s first drawings of the microscopic brain structures \cite{RN189}, neurons have monopolized brain research and emerging disciplines such as computational neuroscience and neuromorphic computing. Therefore, it comes as a surprise to many to learn that up to 90 percent of the brain’s cells are not neurons, but are instead glial cells. The impressive empirical evidence of the importance of non-neuronal cells, particularly astrocytes, in all facets of cognitive processes \cite{ZN49, ZN48, ZN47}, including learning and memory \cite{ZN50, ZN45, ZN52, ZN46, ZN51}, is shaping a paradigm shift where brain function and dysfunction are now seen as phenomena emerging from the interaction between neurons and astrocytes \cite{ZN44, ZN54, ZN55, ZN53}. This also opens prospects for establishing new connections between biological and artificial intelligence at the cellular, the most fundamental level of computing.

Astrocytes receive input \textit{from} neurons and also provide input \textit{to} them. They do this through processes extending from their somas, which can contact thousands nearby synapses \cite{halassa2007synaptic, polykretis2018}. This three body arrangement (astrocyte process, pre-synaptic neuron, post-synaptic neuron) is named a \textit{tripartite synapse}  \cite{araque1999tripartite}. The main signaling mechanism that astrocytes have is the elevation of their \(Ca^{2+}\) concentration \cite{RN51}. The cells communicate with one another by propagating long-distance intercellular waves of \(Ca^{2+}\) ions \cite{RN25}; they are also known to propagate intracellular waves within themselves--the conditions under which these two signaling modes occur remain largely mysterious \cite{RN51}. Individual astrocyte processes respond to pre-synaptic input also with an elevation in their internal \(Ca^{2+}\) levels \cite{RN51}. Interestingly, this neuronal-astrocytic interaction is dynamic and plastic, although little is known about the exact form of this plasticity \cite{RN216,polykretis2018astrocytic}. The timescale of astrocyte \(Ca^{2+}\) excitability was previously believed to be entirely on the order of seconds to hours, yet recent experiments have found a faster astrocytic response to synaptic activity—on the order of hundreds of milliseconds, taking place at the astrocytic process \cite{RN190,polykretis2019}. Although initially liked to hemodynamic response \cite{RN190}, the role of this recently discovered ``fast'' \(Ca^{2+}\) signal remains elusive. 

Astrocytes have long been implicated in learning \cite{han2013} and, recently, they have been deemed as not only necessary but also sufficient cells for new memory formation \cite{adamsky2018astrocytic}. The leading hypothesis—which faces interesting competitors \cite{RN214}—is that memories are stored as the connection strengths between neurons, and these strengths are modified as the organism learns and grows \cite{RN207}. A computationally elegant model of memory, the Hopfield network, incorporates the above features to perform autoassociation: the tag to retrieve a network state is a corrupted version of the state itself \cite{RN212}. Learning in a Hopfield network \cite{RN213} can be understood as creating new attractors in the configuration space of the system, so that when the system is put into a configuration close to one that has been stored (i.e when the network is presented a noisy version of a fundamental memory), it dynamically relaxes towards the nearest fundamental memory, and stay there indefinitely. This model—as well as its continuous analogs \cite{RN215}—have been used to explain neuronal dynamics in several brain regions, including persistent activity in cortex \cite{RN196,RN197} and path integration in hippocampus \cite{RN195}. A challenge for Hopfield-type neural networks is explaining the origin of temporal sequences; in other words, how can a network be constructed so that the tag to retrieve a given sequence of memories is the first memory in that sequence? Hopfield himself recognized this problem and proposed a modification to his original model which allowed for the recall of temporal sequences by using an asymmetric synaptic weight matrix \cite{RN213}. However, this method suffered from instabilities and was difficult to control. Sompolinsky et al. \cite{RN206}, independently and in parallel with Kleinfeld \cite{RN205}, showed that this scheme could be made robustly stable by the introduction of ``slow-synapses''—synapses which compute a weighted average of the pre-synaptic neuron state. 

In this paper, we present a theoretical abstraction of the astrocytic response to neuronal activity, analyze the associated dynamics of the neuron-astrocyte interaction, and derive a computational framework for sequence learning that can be used in SNN algorithms. Specifically, we propose a Hopfield-type recurrent neuronal-astrocytic network, where each synapse is enseathed by an astrocytic process (Fig 1). The network used its neuronal component to learn distinct memories and its astrocytic component to transition between the stored memories. We also suggest a Hebbian-type astrocytic mechanism to learn the transition between stored memories, upon triggering the neuronal network state changes. We validated our model by studying its performance as a function of astrocytic atrophy, following studies on cognition-impairing diseases \cite{verkhratsky2010astrocytes}. Interestingly enough, we found a significant correlation between the level of atrophy and the degree of cognitive impairment, as measured by the error in the associative network's ability to recall a sequence, in agreement with experimental results.

\begin{figure}[t]
\centering
\includegraphics[scale=.041]{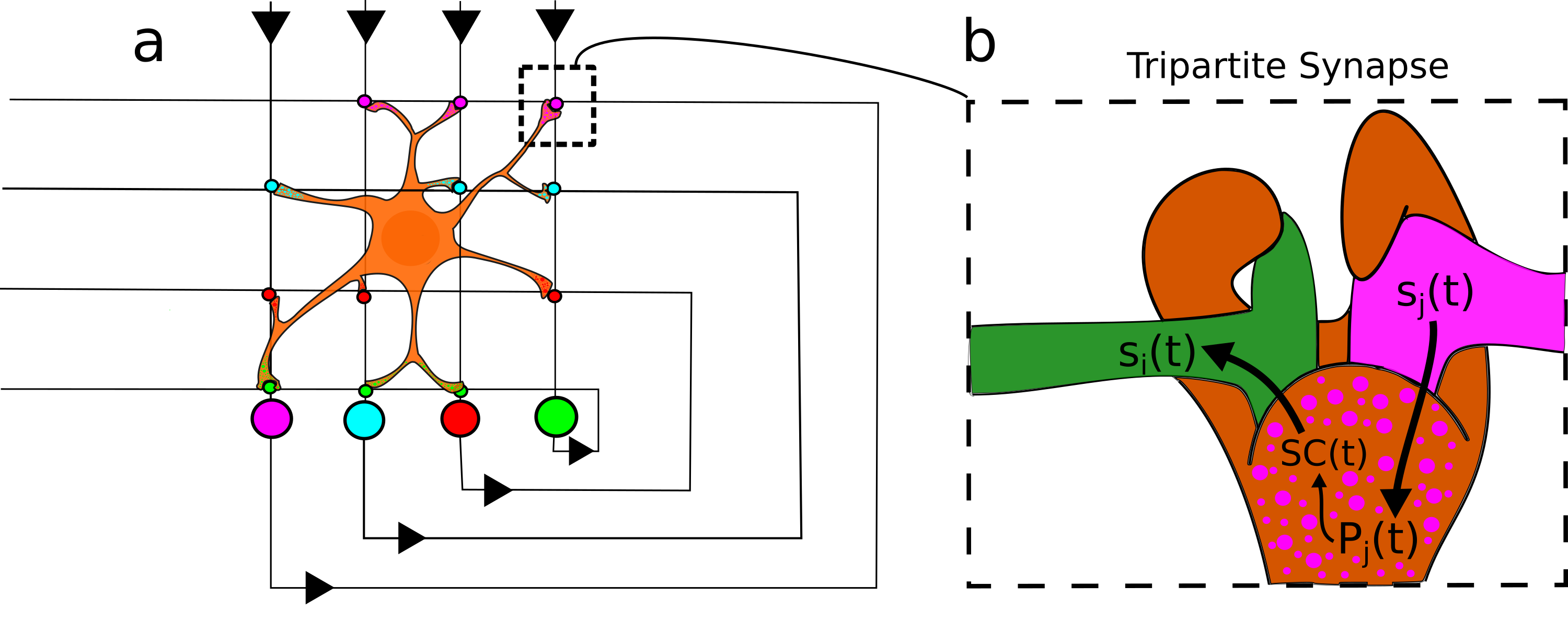}
\caption{ a) An astrocyte ensheathing with its processes a fully-connected, recurrent neuronal network (here, N = 4). Neurons are represented as large colored circles, synaptic connections are represented
as small colored circles, with the color corresponding to the color of the afferent neuron. No self-connections were allowed. The astrocyte processes were independent, and responded only to the presynaptic firing activity at the tripartite synapse. b) The tripartite synapse, showing the flowchart of
a presynaptic activity $s_{j}(t)$ driving the astrocytic process state $P_{j}(t)$ which triggers the SC signal to get injected into the postsynaptic neuron for SC production.}

\end{figure}

\section{Methods}

\subsection{Neurophysiological Background}
Astrocytes share the same mechanisms with neurons as they, too, modulate the flux of ions into and out of the neurons, through gliotransmitters. The current injected into the neurons can be positive or negative--denoted slow-inward current (SIC) and slow-outward current (SOC), respectively \cite{RN203,RN113,RN200}. SICs appear to be released into postsynaptic neurons when the \(Ca^{2+}\) level inside the astrocyte reaches a certain threshold from below \cite{RN112}. Less is known about SOCs, although they seem to follow a similar time-course to SICs \cite{pal2015astrocytic}. 

Here, we present a biophysically plausible model of how astrocytes may employ the SICs and SOCs to enable the transition between memories in a network. We also propose a simple, co-occurence (Hebbian-type) learning rule between the astrocyte process and the post-synaptic neuron, which formalizes the notion of astrocyte-neuron plasticity \cite{RN216}. Incorporating the recently discovered short response time of the astrocytic processes to the presynaptic activity, the memory model uses the astrocytes to trigger the transitions between learned states, where the timing of the transitions was governed by the dynamics of the \(Ca^{2+}\)-dependent SICs and SOCs.

\subsection{Deriving Network Dynamics}

We modeled neurons as zero-temperature, spin-glass units, with 1 and 0 representing the active and quiescent states, respectively. The output of neuron \textit{i} was aligned with the local field, $h_{i}$:
\begin{equation}
s_{i}(t+1)=sgn(h_{i}(t)).
\end{equation}

\begin{figure}[t]
\includegraphics[scale=.145]{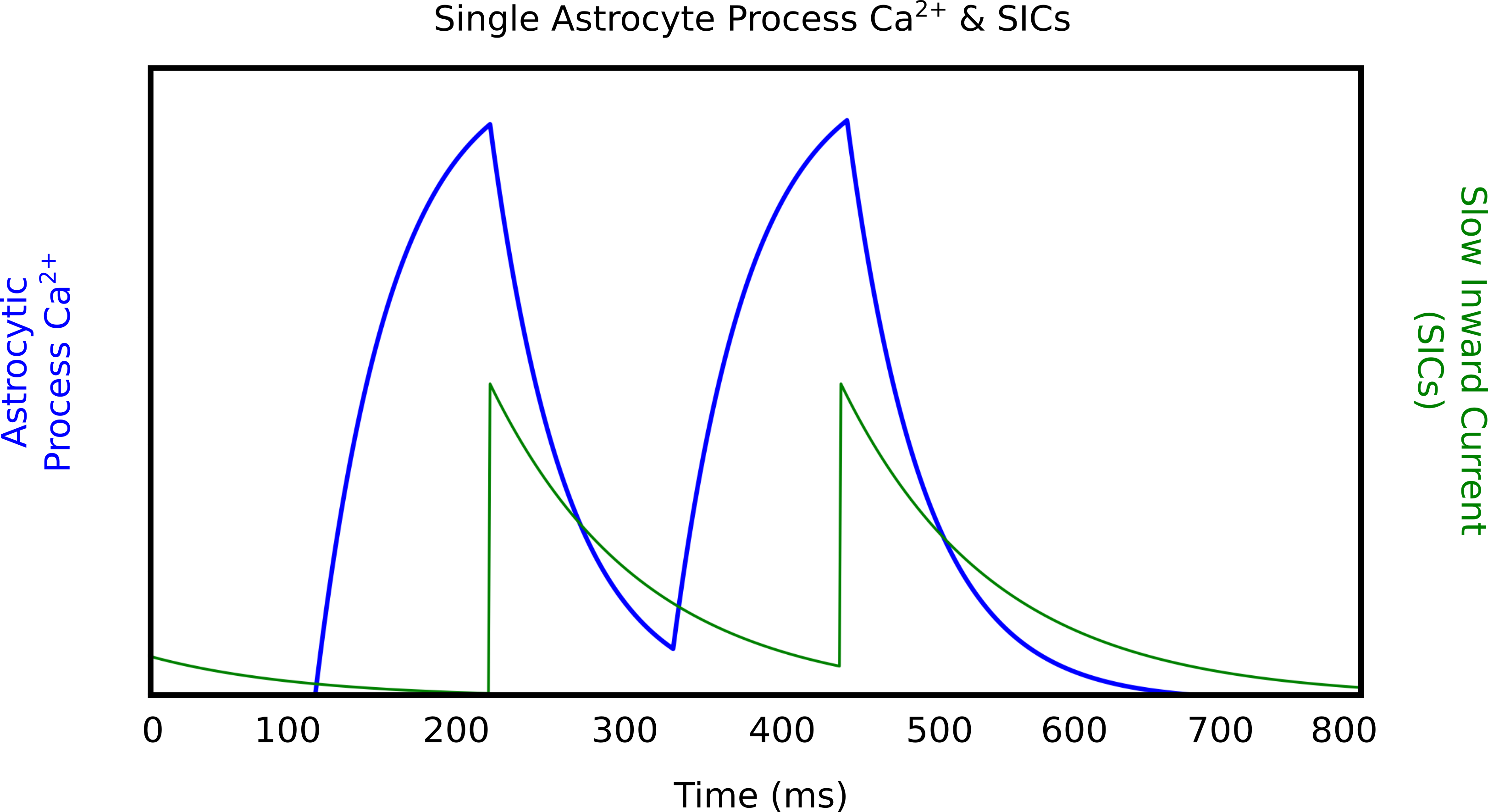}
\caption{The simulated dynamics of the two signals of interest in a single astrocyte process: Local \(Ca^{2+}\) wave (blue) which rises in response to the presynaptic activity and the related SIC (green) which is injected into the postsynaptic neuron. The y-axes are scaled for visibility/clarity, as their relative amplitudes are of no biological relevance for this study.}

\label{fig:2}
\end{figure}

Here, we expanded $h_i$ to include the effects of astrocyte-mediated post-synaptic SICs and SOCs, as follows:

\begin{equation}
h_{i}(t)=h_{i}(t)^{neural}+h_{i}(t)^{astro},
\end{equation}

\begin{equation}
h_{i}(t)^{neural}=\sum_{j=1}^{N}J_{ij}s_{j}(t),
\end{equation}

\begin{equation}
h_{i}(t)^{astro}=\sum_{j=1}^{N}T_{ij}SC_{j}(t),
\end{equation}

where \(N\) was the number of neurons, $J_{ij}$ was the stabilizing, symmetric matrix, and \(T_{ij}\) was the matrix of amplitudes for the astrocyte-mediated slow-currents (SCs), either a SIC or a SOC. Following experimental studies suggesting that most synapses are enseathed by astrocytic processes \cite{bernardinelli2014astrocyte}, all \(N^2\) synapses in the network were tripartite synapses. Since all the processes that take neuron \(i\) as its input were synchronized, the vector of SCs was of size \(N^2/N\). Let \(\xi_{i}^{\mu}\)denote the activity of neuron \textit{i} during memory \(\mu\), and \(m\) denote the number of memories stored in the network:

\begin{equation}
J_{ij}=\frac{1}{N}\sum_{\mu}^{m}(2\xi_{i}^{\mu}-1)(2\xi_{j}^{\mu}-1),i\neq j,
\end{equation}

\begin{equation}
T_{ij}=\frac{\lambda}{N}\sum_{\mu}^{q}(2\xi_{i}^{\mu+1}-1)(2\xi_{j}^{\mu}-1),i\neq j,
\end{equation}

where \(q<m\), the \(\xi_{i}^{\mu+1}\xi_{j}^{\mu}\) terms define the sequence of memories, and \(\lambda\) controls the relative strength between the two matrices. We set all diagonal elements of both matrices to zero, not allowing self-connections. Following experimental evidence \cite{pal2015astrocytic}, the SCs exponentially decayed (see Fig \ref{fig:2}) after a rapid rise time (negligible). :

\begin{equation}
SC=e^{\frac{t-\delta_{cal}}{\tau_{SC}}},
\end{equation}

where \(\delta_{cal}\) is the time at which the astrocyte \(Ca^{2+}\) reached the SC-release threshold, \(c_{thresh}\). We propose a minimal model for \(Ca^{2+}\) level in the astrocyte process, \(P_j\), where the time evolution of \(P_j\) depends linearly on the activity of the presynaptic neuron \(s_j\) and the previous state. Dropping the \(j\) subscript, an astrocyte process at time t+1, denoted \(P_{t+1}\) is given by:

\begin{equation}
P_{t+1}=\alpha P_{t}+\beta s_{t}, 
\end{equation}

where \(0\leq\alpha<1\). 

\subsection{Modeling the Effects of Astrocytic Atrophy to Memory Recall}

To evaluate our model, we randomly selected a percentage (from 0 to 100\%) of astrocytic processes that were atrophied. Specifically, for each one of the selected processes, we introduced a gain (from 0 to 1, representing high and no atrophy, respectively). 

We validated the network performance as follows: for an ordered sequence of q memories, the performance error was the number of times a memory did not appear in its appropriate spot, divided by the number of possible errors (to ensure the error is between 0 and 1). We excluded the first memory from the evaluation, as it did not depend on the astrocyte dynamics.

\section{Results}

\subsection{Transition Times and Stability}

The above equation can be solved in terms of \(s_t\), \(\alpha\) and \(\beta\) alone by defining the operator  \(\hat{L}\) such that:
\[\hat{L}P_{t}\equiv P_{t-1}\]
\[\hat{L}^{2}P_{t}\equiv P_{t-2}\]
We can then arrive at an expression for \(P_t\):
\[P_{t}=\frac{\beta s_{t}}{1-\alpha\hat{L}}=\beta\sum_{t'=0}^{\infty}(\alpha\hat{L})^{t'}s_{t}=\beta\sum_{t'=0}^{\infty}\alpha^{t'}s_{t-t'}\].
 
\begin{figure}[t]
\includegraphics[scale=.127]{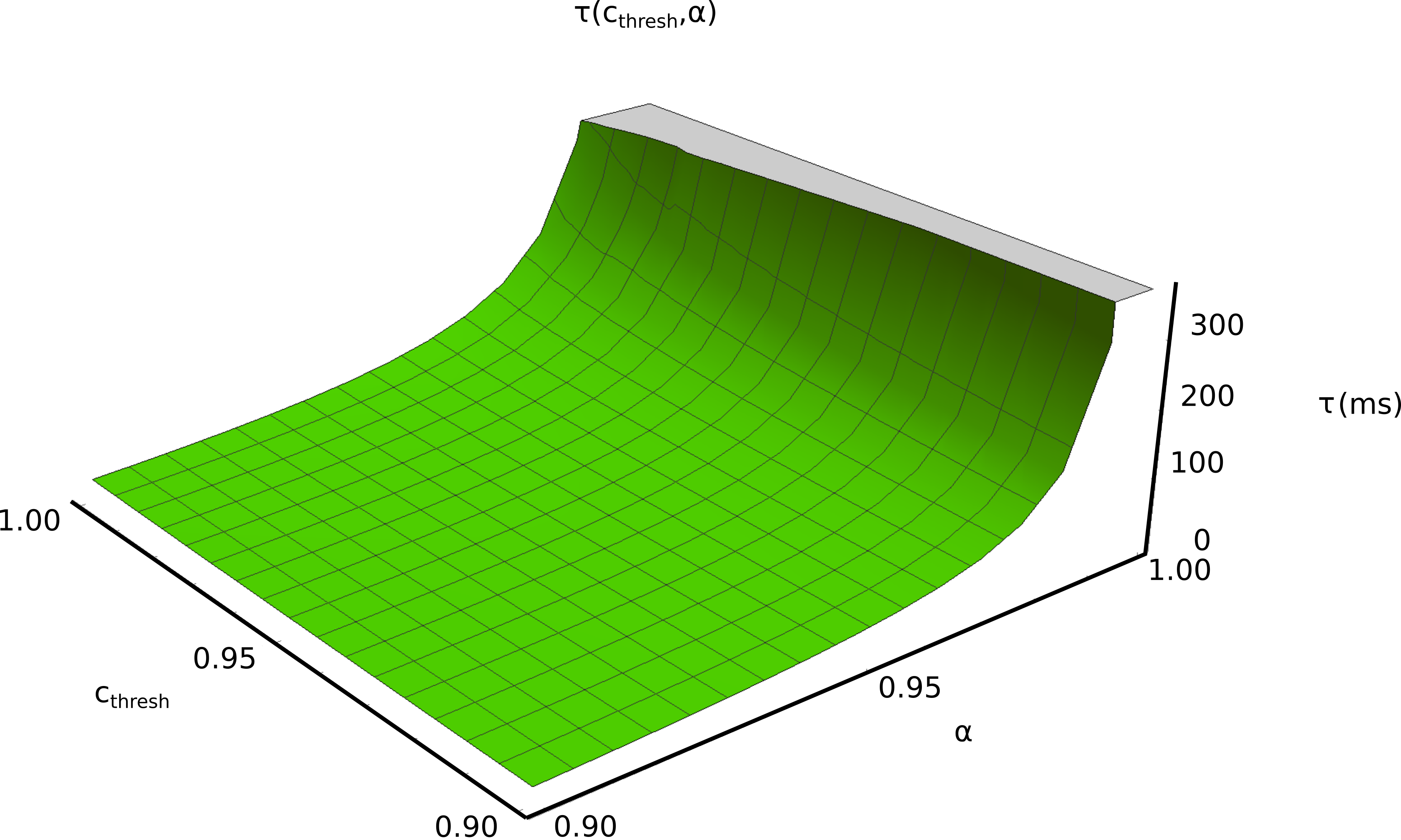}
\caption{The derived amount of time the network spends in each quasi-attractor \(\tau\), as a function of \(\alpha\) and \(c_{thresh}\). For a fixed \(\alpha\), increasing \(c_{thresh}\) increases the time it takes for an astrocyte processes to release a SC, thus increasing \(\tau\).}

\label{fig:3}
\end{figure}

We can determine the amount of time it takes for the \(Ca^{2+}\) to reach the SC-release threshold, which in turn will determine the duration of time a network spends in a given quasi-attractor (\(\tau\)) (see Fig \ref{fig:3}). The analysis is simplified in the continuous limit:

\begin{equation}
\frac{c_{thresh}}{\beta}=\intop_{0}^{\tau}\alpha^{\tau-'t}dt',   
\end{equation}

which has the general solution:

 \begin{equation}
\tau=\frac{ln(\frac{c_{thresh}}{\beta}ln(\alpha)+1)}{ln(\alpha)}.
\end{equation}

Though the choice of \(\beta\) and \(c_{thresh}\) are arbitrary (provided that \(\tau\) remains a positive real number and \(0<c_{thresh} <1\)), if one makes the simplifying assumption that \(\beta\alpha^{t}\) is normalized to unity, this expression becomes:

\begin{equation}
\tau=\frac{ln(1-c_{thresh})}{ln(\alpha)}.
\end{equation}

This equation says that for a fixed \(\alpha\), the time to reach the SC-threshold scales logarithmically with the threshold value. The biological interpretation of the normalization is that the more the astrocyte processes depends on its own \(Ca^{2+}\) level in the previous timestep (determined by \(\alpha\)), the less it depends on the presynaptic neuronal activity (determined by \(\beta\)), since normalization implies that \(\beta=ln(\frac{1}{\alpha})\). 

Having this parameter, we can now examine the dynamics of the network in detail. The analysis is simplified by switching to the \(s_i = \pm 1\) neuronal representation, which is related to the \(s_i = 0,1\) representation by the transformation \(2s_i -1\).

Imagine that at time \(t = 0\) the network has entered into the attractor for memory \(\xi^{1}\). The \(Ca^{2+}\) thresholds have not been hit (i.e. \(SC_{j}(t) = 0\) for all j).Therefore, the total field felt by neuron i is simply:

\begin{equation}
h_{i}(t)=\sum_{j=1}^{N}J_{ij}\xi_{i}^{1}=\xi_{i}^{1}+\textit{noise}.
\end{equation}

If we assume low loading (i.e \(p<<N\)), the noise term vanishes. This field persists until \(t = \tau\), the time at which \(c_{thresh}\) is reached by the active astrocyte processes. Now the field becomes:

\begin{equation}
h_{i}(t)=\frac{1}{N}\sum_{\mu=1}^{m}\xi_{i}^{\mu}\xi_{j}^{\mu}\xi_{j}^{1}+\frac{\lambda}{N}\sum_{\mu=1}^{q}\xi_{i}^{\mu+1}\xi_{j}^{\mu}SC_{j}(t).  
\end{equation}

\begin{figure}[t]
\includegraphics[scale=.16]{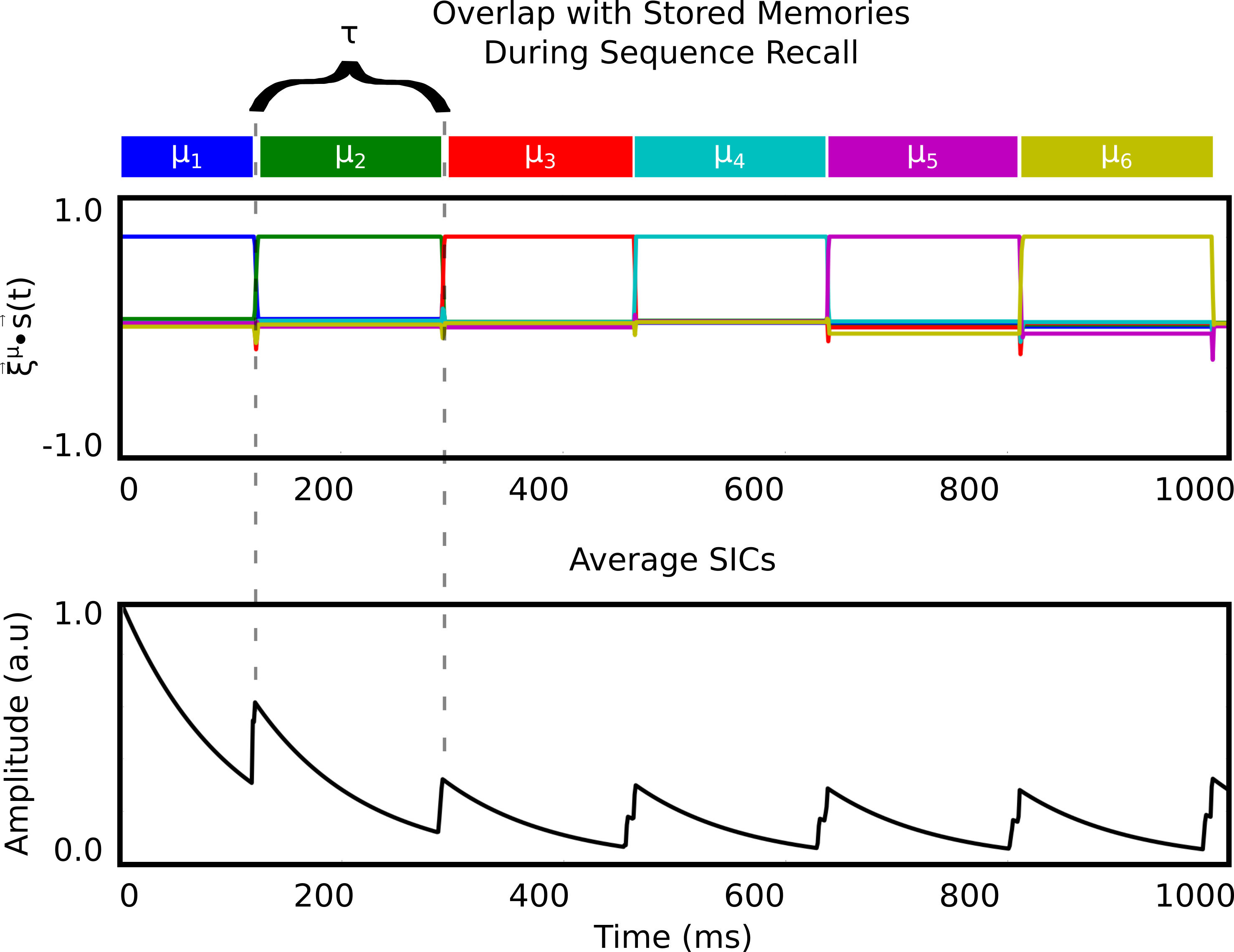}
\caption{Network demonstration: The overlap of the neuronal network state with the stored memories. N = 500, p=7, q=6.  Lower inset: average SCs released into the post-synaptic neurons. The slight divergence in SC timing is due to the asynchronous dynamics of the network during state transitions.}
\label{fig:4}
\end{figure}

Since the SCs are only released from an astrocyte process if the neuron has been in the active state for \(0<t<\tau\), the vector of SCs at \(t=\tau\) is equal to the vector of neuron states when \(0<t<\tau\). In other words, we identify \(SC_{j}(t) = \xi_{j}^1\), which permits the simplification:

\begin{equation}
h_{i}(t)=\xi_{i}^{1}+\lambda\xi_{i}^{2}.
\end{equation}
In the zero noise limit, the neurons will align with memory \(\xi^{2}\). This field persists until \(t = 2\tau\), when the next transition is precipitated by the astrocyte processes (see Fig \ref{fig:4}). 

\subsection{Model Generality}

An important point to note is that the results of the model are insensitive to the choice of response function for astrocytic \(Ca^{2+}\), so long as the \(Ca^{2+}\) crosses the threshold with some periodicity. It is interesting to consider cases when $\tau$ is time-dependent, since simulations of biophysically-detailed \(Ca^{2+}\) response \cite{RN25} suggest that astrocytes can perform FM, and AFM encoding of synaptic information. Let us consider the simple case of a frequency modulated sinusoid and its first time derivative, which can be written in the following way:

\[y(t)=cos(\omega(t)t)\]
 
 and 
\[\frac{dy}{dt}=[\frac{d\omega}{dt}t+\omega(t)]sin(\omega(t)t)\]

To solve for \(\tau\), we attempt to solve for t such that: \(y(t_{thresh})=0\)  and \(\frac{dy}{dt}|_{t_{thresh}}>0\). Where we assume that SC-threshold equals zero without loss of generality. 

Since the SC-threhsold must be crossed from below for the astrocyte to release gliotransmitters. For example, if \(\omega(t)=w_{0}t\)  (i.e the frequency increases linearly with time) then one can easily show that the time between the \(n^{th}\) (\(n=1,2,3,4...\)) SC-threshold crossings (i.e \(\tau\)
 ) can be written: 
\[\tau_{n}=\sqrt{\frac{\pi}{2\omega_{0}}}(\sqrt{4n+1}-\sqrt{4n-3})\]
which, for large \(n\), approximately equals \(\sqrt{\frac{\pi}{2\omega_{0}n}}\). Note that \(\tau_{n}\)  tends to zero for large \(n\) , as one would expect when the frequency tends to infinity.

\subsection{Astrocytic Learning}

For learning, we propose a simple, Hebbian-type mechanism by which the neuronal-astrocytic network could arrive at the correct form of the matrix T. Assume that at \(t = 0\) the network is presented a pattern, \(\xi^{\mu}\), until some later time \(t = t_{switch}\) when the network is presented \(\xi^{\mu+1}\). If \(t_{switch}\) >>0, the astrocyte process which takes neuron \(i\) as its input will be very nearly equal to \(\xi^{\mu}\). At \(t = t_{switch}\), the astrocyte process correlates its current state with the state of the post-synaptic neuron and adjusts the levels of future gliotransmitter release accordingly—changing the sign and amplitude of future SC release. This can be expressed as follow:

\begin{equation}
\Delta T_{ij}=\eta s_{i}(t_{switch})P_{j}(t_{switch})=\eta s_{i}(t_{switch})s_{j}(0)=\eta\xi_{i}^{\mu+1}\xi_{i}^{\mu},
\end{equation}

which yields exactly the T-matrix used above (assuming the learning rate, \(\eta\), equals 1 and the sequence is presented to the network exactly one time), in the \(s_i =\pm 1\) representation. Interestingly enough, this mechanism requires retrograde signaling between the post-synaptic neuron and astrocyte process, which is known to occur through endocannabinoid mediated pathways \cite{RN200}.

\subsection{Evaluation: Memory Retrieval Robustness to Astrocytic Impairment}

We show that the degree of a cognitive impairment in our model depended strongly on the degree of astrocytic atrophy (Fig. 5), in agreement with experimental data \cite{verkhratsky2010astrocytes,chung2015glia,ZN44} . This result can be understood on the basis of stability arguments. Atrophying the astrocyte SC signal is equivalent to decreasing $\lambda$ (see equation 6). By inspection of equation 14, one can see that increasing the effective $\lambda$ below 1 at a given neuron will make that neuron unstable. After a critical point of unstable neurons was reached, the network error rapidly increased to 1.

\begin{figure}[t]
\includegraphics[scale=.4]{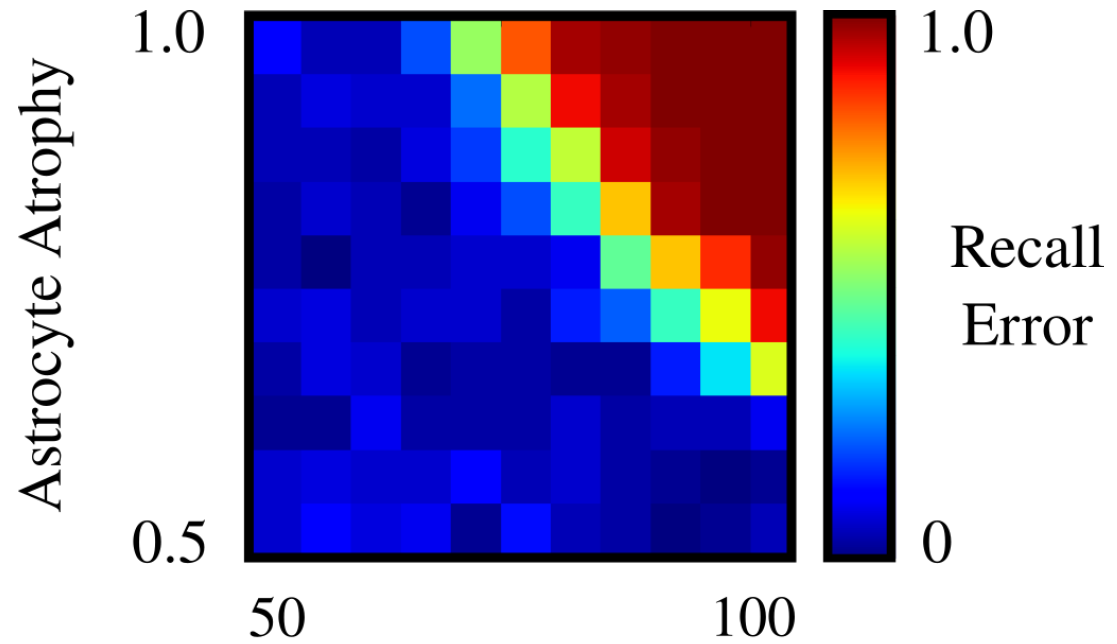}
\caption{Network evaluation: The performance of the network as function of increasing astrocytic atrophy. As the fraction of affected astrocytes (x-axis) and the degree of atrophy (y-axis) increase, the error in sequence recall also increases. The simulation was run 50 times per $(x,y)$ point on the error map, the error shown is the average error
of those 50 trials.}
\label{fig:5}
\end{figure}

\section{Discussion}

Here, we present a neuronal-astrocytic Hopfield-type network and mathematically derive the dynamics of the interactions between neurons and astrocytes to effectively transition between memories. Building from the bottom-up a computational role for astrocytes, our work got inspired by the last decade of memory-related glial research and the recent studies on the fast signaling taking place between astrocytes and neurons; it was also qualitatively evaluated by studies on memory impairment. We demonstrated how astrocytes were sufficient to trigger transitions between stored memories. By injecting a ``fast'' \(Ca^{2+}\) triggered current in the postsynaptic neuron, astrocytes modulated the neuronal activity into coherent, predictable patterns across time, sharpening a particular input and, thereby, recalling a learned memory sequence. This ability of astrocytes to modulate neuronal excitability and synaptic strength can have several implications, both theoretical and practical, for neuromorphic algorithms. 

On the theory side, Sompolinsky et al showed mathematically that the introduction of "slow-synapses"—synapses that perform a running average of pre-synaptic input using a weighting function $w(t-\tau)$ would
stabilize the sequence of memories \cite{RN206}. The authors were able to show this quite generally, placing only a few requirements on the choice of $w(t-\tau)$. Here, we showed how the time delay,$\tau$, emerges naturally out of the dynamics of \(Ca^{2+}\) dependent gliotransmission, as constrained by experimental data. Formally, the dynamics of our biomimetic approach are mathematically equivalent to the case where: $w(t-\tau)=\delta(t-\tau)$, where $\delta(t)$ is the delta function. Interestingly enough, these results demonstrate how biologically plausible models of cellular processes identified in recent brain studies, may provide a mechanistic explanation for theoretical analyses conducted at the network scale, decades ago.

Previous astrocyte modelling efforts have focused primarily on reproducing the \(Ca^{2+}\) response of astrocytes by numerically solving systems of coupled differential equations, where each equation determines the time evolution of an organelle believed to be important for the mechanism of \(Ca^{2+}\)
oscillation, such as ATP or IP3. While crucial for our understanding of astrocytic Ca2+, these studies typically shy away from proposing and modelling actual computational roles for astrocytic function. A notable exception is from Wade et al. \cite{RN112}, who showed that astrocyte oscillations can induce synchrony in unconnected neurons, using the same mechanism of Ca2+ dependent gliotransmission as our study. De Pitta et al.\cite{de2016astrocytes} also explored the role of astrocyte \(Ca^{2+}\) oscillations in long term potentiation (LTP) and long term depression (LTD), two phenomena known to play a key role in brain computation and learning. While previous efforts on fleshing out mechanisms known to be involved in brain computation, our work presents an end-to-end solution, an associative network that uses astrocytic mechanisms to perform a function, sequence memory recall.

On the applications side, by enabling the most faithful representation of neurons, networks and brain systems, neuromorphic computing allows for studies that not necessarily follow a mainstream machine learning direction. For example, neuronal-astrocytic networks on neuromorphic chips may be used to study hypotheses on astrocytes failing to perform their critical synaptic functions, as we did here. Mounting evidence suggests that astrocytes change their strength of their connections in learning \cite{RN192}. Crucially, little is known about the interactions between the localized \(Ca^{2+}\) response of individual astrocyte processes and the global \(Ca^{2+}\) events \cite{RN190}. We speculate that this interaction might enable training the network to not only learn the correct sequence of memories, but also the time spent in each memory for a given sequence. This is biologically faithful, as the amount of time spent in each memory (e.g the duration of a note when humming a melody) is crucial for correctly recalling a learned sequence. In the framework of our model, this can be achieved by dynamically modifying \(\tau\)--which in turn is controlled by the SC-release threshold and the astrocytic sensitivity to pre-synaptic activity. Studies like this might be enabled through careful coupling of computational neuroscience and neuromorphic hardware, using, e.g., specialized astrocytic modules like the one we presented recently \cite{RN_34}. 

Learning is at the core of neuromorphic computing. By reproducing the functional organization of neuronal-astrocytic networks, as well as the dynamics of astrocytic \(Ca^{2+}\) activity and astrocyte-neuron interactions, we computationally suggested a learning role for astrocytes operating on temporal and spatial scales that are larger than the ones in neurons. The underlying mechanisms of having parallel processing on different temporal and spatial scales is an open question in brain science, but it is already highly regarded as a computational method that increases the processing efficiency of a system: Our work tackles this problem by combining millisecond-scale neuronal activity with the comparatively slow \(Ca^{2+}\) activity of astrocytes. 

Most of the neuroscience knowledge accumulated over the past couple decades has yet to be funnelled in artificial intelligence. Being shadowed by the wide applications of neural nets, we might not appreciate that the fast accumulating knowledge on the biological principles of intelligence is only partially explored, or exploited, on the computational side. Can we establish new push-pull dynamics between newly identified biological principles of intelligence and the computational primitives used to build our artificial models of brain computation? To explore this fascinating possibility, our work couples computational modeling and neuromorphic computing to introduce to neurocomputing a long-neglected non-neuronal cell, astrocytes, which are now placed alongside neurons, as key cells for learning. The results present here suggest that the addition of astrocytes as a second processing unit to neuromorphic chips is a direction worth pursuing. 

\section{Conclusion}
The emergence of neuromorphic computing calls for a bottom-up rethinking of computational algorithms that can seamlessly integrate into non-Von Neumann hardware, promising unparalleled energy-efficiency and a robust yet versatile alternative to the brittle inference-based artificial intelligence solutions. The work we presented here aims to bring us closer to realize this promise by building a theoretical framework that supports a functional role for astrocytes in associative networks. The further scaling of the astrocytic roles will support real-world neuromorphic applications, where astrocytes will be able to mine intrinsically noisy data, by virtue of their low spatial and temporal resolution. Drawing from newly identified primitives of biological intelligence, this paper paves the way for a "built-in" versatility and robustness in intelligent neuromorphic systems.

\bibliographystyle{ACM-Reference-Format}
\bibliography{sample-base}


\begin{thebibliography}{88}


\ifx \showCODEN    \undefined \def \showCODEN     #1{\unskip}     \fi
\ifx \showDOI      \undefined \def \showDOI       #1{#1}\fi
\ifx \showISBNx    \undefined \def \showISBNx     #1{\unskip}     \fi
\ifx \showISBNxiii \undefined \def \showISBNxiii  #1{\unskip}     \fi
\ifx \showISSN     \undefined \def \showISSN      #1{\unskip}     \fi
\ifx \showLCCN     \undefined \def \showLCCN      #1{\unskip}     \fi
\ifx \shownote     \undefined \def \shownote      #1{#1}          \fi
\ifx \showarticletitle \undefined \def \showarticletitle #1{#1}   \fi
\ifx \showURL      \undefined \def \showURL       {\relax}        \fi
\providecommand\bibfield[2]{#2}
\providecommand\bibinfo[2]{#2}
\providecommand\natexlab[1]{#1}
\providecommand\showeprint[2][]{arXiv:#2}

\bibitem[\protect\citeauthoryear{Adamsky, Kol, Kreisel, Doron, Ozeri-Engelhard,
  Melcer, Refaeli, Horn, Regev, and Groysman}{Adamsky et~al\mbox{.}}{2018a}]%
        {ZN50}
\bibfield{author}{\bibinfo{person}{Adar Adamsky}, \bibinfo{person}{Adi Kol},
  \bibinfo{person}{Tirzah Kreisel}, \bibinfo{person}{Adi Doron},
  \bibinfo{person}{Nofar Ozeri-Engelhard}, \bibinfo{person}{Talia Melcer},
  \bibinfo{person}{Ron Refaeli}, \bibinfo{person}{Henrike Horn},
  \bibinfo{person}{Limor Regev}, {and} \bibinfo{person}{Maya Groysman}.}
  \bibinfo{year}{2018}\natexlab{a}.
\newblock \showarticletitle{Astrocytic activation generates de novo neuronal
  potentiation and memory enhancement}.
\newblock \bibinfo{journal}{\emph{Cell}} \bibinfo{volume}{174},
  \bibinfo{number}{1} (\bibinfo{year}{2018}), \bibinfo{pages}{59--71. e14}.
\newblock
\showISSN{0092-8674}


\bibitem[\protect\citeauthoryear{Adamsky, Kol, Kreisel, Doron, Ozeri-Engelhard,
  Melcer, Refaeli, Horn, Regev, Groysman, et~al\mbox{.}}{Adamsky
  et~al\mbox{.}}{2018b}]%
        {adamsky2018astrocytic}
\bibfield{author}{\bibinfo{person}{Adar Adamsky}, \bibinfo{person}{Adi Kol},
  \bibinfo{person}{Tirzah Kreisel}, \bibinfo{person}{Adi Doron},
  \bibinfo{person}{Nofar Ozeri-Engelhard}, \bibinfo{person}{Talia Melcer},
  \bibinfo{person}{Ron Refaeli}, \bibinfo{person}{Henrike Horn},
  \bibinfo{person}{Limor Regev}, \bibinfo{person}{Maya Groysman},
  {et~al\mbox{.}}} \bibinfo{year}{2018}\natexlab{b}.
\newblock \showarticletitle{Astrocytic activation generates de novo neuronal
  potentiation and memory enhancement}.
\newblock \bibinfo{journal}{\emph{Cell}} \bibinfo{volume}{174},
  \bibinfo{number}{1} (\bibinfo{year}{2018}), \bibinfo{pages}{59--71}.
\newblock


\bibitem[\protect\citeauthoryear{Alcorn, Li, Gong, Wang, Mai, Ku, and
  Nguyen}{Alcorn et~al\mbox{.}}{[n.d.]}]%
        {RN_15}
\bibfield{author}{\bibinfo{person}{Michael~A Alcorn}, \bibinfo{person}{Qi Li},
  \bibinfo{person}{Zhitao Gong}, \bibinfo{person}{Chengfei Wang},
  \bibinfo{person}{Long Mai}, \bibinfo{person}{Wei-Shinn Ku}, {and}
  \bibinfo{person}{Anh Nguyen}.} \bibinfo{year}{[n.d.]}\natexlab{}.
\newblock \showarticletitle{Strike (with) a pose: Neural networks are easily
  fooled by strange poses of familiar objects}. In
  \bibinfo{booktitle}{\emph{Proceedings of the IEEE Conference on Computer
  Vision and Pattern Recognition}}. \bibinfo{pages}{4845--4854}.
\newblock


\bibitem[\protect\citeauthoryear{Araque, Parpura, Sanzgiri, and Haydon}{Araque
  et~al\mbox{.}}{1999}]%
        {araque1999tripartite}
\bibfield{author}{\bibinfo{person}{Alfonso Araque}, \bibinfo{person}{Vladimir
  Parpura}, \bibinfo{person}{Rita~P Sanzgiri}, {and} \bibinfo{person}{Philip~G
  Haydon}.} \bibinfo{year}{1999}\natexlab{}.
\newblock \showarticletitle{Tripartite synapses: glia, the unacknowledged
  partner}.
\newblock \bibinfo{journal}{\emph{Trends in neurosciences}}
  \bibinfo{volume}{22}, \bibinfo{number}{5} (\bibinfo{year}{1999}),
  \bibinfo{pages}{208--215}.
\newblock


\bibitem[\protect\citeauthoryear{Barres}{Barres}{2008}]%
        {ZN44}
\bibfield{author}{\bibinfo{person}{Ben~A Barres}.}
  \bibinfo{year}{2008}\natexlab{}.
\newblock \showarticletitle{The mystery and magic of glia: a perspective on
  their roles in health and disease}.
\newblock \bibinfo{journal}{\emph{Neuron}} \bibinfo{volume}{60},
  \bibinfo{number}{3} (\bibinfo{year}{2008}), \bibinfo{pages}{430--440}.
\newblock
\showISSN{0896-6273}


\bibitem[\protect\citeauthoryear{Bazargani and Attwell}{Bazargani and
  Attwell}{2016}]%
        {RN51}
\bibfield{author}{\bibinfo{person}{Narges Bazargani} {and}
  \bibinfo{person}{David Attwell}.} \bibinfo{year}{2016}\natexlab{}.
\newblock \showarticletitle{Astrocyte calcium signaling: the third wave}.
\newblock \bibinfo{journal}{\emph{Nature neuroscience}} \bibinfo{volume}{19},
  \bibinfo{number}{2} (\bibinfo{year}{2016}), \bibinfo{pages}{182--189}.
\newblock
\showISSN{1097-6256}


\bibitem[\protect\citeauthoryear{Benarroch}{Benarroch}{2005}]%
        {ZN54}
\bibfield{author}{\bibinfo{person}{Eduardo~E. Benarroch}.}
  \bibinfo{year}{2005}\natexlab{}.
\newblock \showarticletitle{Neuron-Astrocyte Interactions: Partnership for
  Normal Function and Disease in the Central Nervous System}.
\newblock \bibinfo{journal}{\emph{Mayo Clinic Proceedings}}
  \bibinfo{volume}{80}, \bibinfo{number}{10} (\bibinfo{year}{2005}),
  \bibinfo{pages}{1326--1338}.
\newblock
\showISSN{0025-6196}
\urldef\tempurl%
\url{http://www.sciencedirect.com/science/article/pii/S0025619611617606}
\showURL{%
\tempurl}


\bibitem[\protect\citeauthoryear{Benjamin, Gao, McQuinn, Choudhary,
  Chandrasekaran, Bussat, Alvarez-Icaza, Arthur, Merolla, and Boahen}{Benjamin
  et~al\mbox{.}}{2014}]%
        {RN_22}
\bibfield{author}{\bibinfo{person}{Ben~Varkey Benjamin},
  \bibinfo{person}{Peiran Gao}, \bibinfo{person}{Emmett McQuinn},
  \bibinfo{person}{Swadesh Choudhary}, \bibinfo{person}{Anand~R
  Chandrasekaran}, \bibinfo{person}{Jean-Marie Bussat},
  \bibinfo{person}{Rodrigo Alvarez-Icaza}, \bibinfo{person}{John~V Arthur},
  \bibinfo{person}{Paul~A Merolla}, {and} \bibinfo{person}{Kwabena Boahen}.}
  \bibinfo{year}{2014}\natexlab{}.
\newblock \showarticletitle{Neurogrid: A mixed-analog-digital multichip system
  for large-scale neural simulations}.
\newblock \bibinfo{journal}{\emph{Proc. IEEE}} \bibinfo{volume}{102},
  \bibinfo{number}{5} (\bibinfo{year}{2014}), \bibinfo{pages}{699--716}.
\newblock
\showISSN{0018-9219}


\bibitem[\protect\citeauthoryear{Bernardinelli, Muller, and
  Nikonenko}{Bernardinelli et~al\mbox{.}}{2014}]%
        {bernardinelli2014astrocyte}
\bibfield{author}{\bibinfo{person}{Yann Bernardinelli},
  \bibinfo{person}{Dominique Muller}, {and} \bibinfo{person}{Irina Nikonenko}.}
  \bibinfo{year}{2014}\natexlab{}.
\newblock \showarticletitle{Astrocyte-synapse structural plasticity}.
\newblock \bibinfo{journal}{\emph{Neural plasticity}}  \bibinfo{volume}{2014}
  (\bibinfo{year}{2014}).
\newblock


\bibitem[\protect\citeauthoryear{Bing, Meschede, Huang, Chen, Rohrbein, Akl,
  and Knoll}{Bing et~al\mbox{.}}{[n.d.]}]%
        {RN_35}
\bibfield{author}{\bibinfo{person}{Zhenshan Bing}, \bibinfo{person}{Claus
  Meschede}, \bibinfo{person}{Kai Huang}, \bibinfo{person}{Guang Chen},
  \bibinfo{person}{Florian Rohrbein}, \bibinfo{person}{Mahmoud Akl}, {and}
  \bibinfo{person}{Alois Knoll}.} \bibinfo{year}{[n.d.]}\natexlab{}.
\newblock \showarticletitle{End to end learning of spiking neural network based
  on r-stdp for a lane keeping vehicle}. In \bibinfo{booktitle}{\emph{2018 IEEE
  International Conference on Robotics and Automation (ICRA)}}.
  \bibinfo{publisher}{IEEE}, \bibinfo{pages}{1--8}.
\newblock
\showISBNx{1538630818}


\bibitem[\protect\citeauthoryear{Blouw, Choo, Hunsberger, and Eliasmith}{Blouw
  et~al\mbox{.}}{2018}]%
        {RN_33}
\bibfield{author}{\bibinfo{person}{Peter Blouw}, \bibinfo{person}{Xuan Choo},
  \bibinfo{person}{Eric Hunsberger}, {and} \bibinfo{person}{Chris Eliasmith}.}
  \bibinfo{year}{2018}\natexlab{}.
\newblock \showarticletitle{Benchmarking keyword spotting efficiency on
  neuromorphic hardware}.
\newblock \bibinfo{journal}{\emph{arXiv preprint arXiv:1812.01739}}
  (\bibinfo{year}{2018}).
\newblock


\bibitem[\protect\citeauthoryear{Blum, Dietmüller, Milde, Conradt, Indiveri,
  and Sandamirskaya}{Blum et~al\mbox{.}}{[n.d.]}]%
        {RN_36}
\bibfield{author}{\bibinfo{person}{Hermann Blum}, \bibinfo{person}{Alexander
  Dietmüller}, \bibinfo{person}{Moritz~B Milde}, \bibinfo{person}{Jörg
  Conradt}, \bibinfo{person}{Giacomo Indiveri}, {and} \bibinfo{person}{Yulia
  Sandamirskaya}.} \bibinfo{year}{[n.d.]}\natexlab{}.
\newblock \showarticletitle{A neuromorphic controller for a robotic vehicle
  equipped with a dynamic vision sensor}. In
  \bibinfo{booktitle}{\emph{Robotics: Science and Systems}}.
\newblock


\bibitem[\protect\citeauthoryear{Cajal}{Cajal}{1888}]%
        {RN189}
\bibfield{author}{\bibinfo{person}{Santiago Ramón~y Cajal}.}
  \bibinfo{year}{1888}\natexlab{}.
\newblock \bibinfo{booktitle}{\emph{Estructura de los centros nerviosos de las
  aves}}. Vol.~\bibinfo{volume}{1}.
\newblock \bibinfo{address}{Rev. Trim. Histol. Norm. Pat.}
\newblock


\bibitem[\protect\citeauthoryear{Cassidy, Merolla, Arthur, Esser, Jackson,
  Alvarez-Icaza, Datta, Sawada, Wong, and Feldman}{Cassidy
  et~al\mbox{.}}{[n.d.]}]%
        {RN_21}
\bibfield{author}{\bibinfo{person}{Andrew~S Cassidy}, \bibinfo{person}{Paul
  Merolla}, \bibinfo{person}{John~V Arthur}, \bibinfo{person}{Steve~K Esser},
  \bibinfo{person}{Bryan Jackson}, \bibinfo{person}{Rodrigo Alvarez-Icaza},
  \bibinfo{person}{Pallab Datta}, \bibinfo{person}{Jun Sawada},
  \bibinfo{person}{Theodore~M Wong}, {and} \bibinfo{person}{Vitaly Feldman}.}
  \bibinfo{year}{[n.d.]}\natexlab{}.
\newblock \showarticletitle{Cognitive computing building block: A versatile and
  efficient digital neuron model for neurosynaptic cores}. In
  \bibinfo{booktitle}{\emph{The 2013 International Joint Conference on Neural
  Networks (IJCNN)}}. \bibinfo{publisher}{IEEE}, \bibinfo{pages}{1--10}.
\newblock
\showISBNx{1467361291}


\bibitem[\protect\citeauthoryear{Chaudhuri and Fiete}{Chaudhuri and
  Fiete}{2016}]%
        {RN212}
\bibfield{author}{\bibinfo{person}{Rishidev Chaudhuri} {and}
  \bibinfo{person}{Ila Fiete}.} \bibinfo{year}{2016}\natexlab{}.
\newblock \showarticletitle{Computational principles of memory}.
\newblock \bibinfo{journal}{\emph{Nature neuroscience}} \bibinfo{volume}{19},
  \bibinfo{number}{3} (\bibinfo{year}{2016}), \bibinfo{pages}{394--403}.
\newblock
\showISSN{1097-6256}


\bibitem[\protect\citeauthoryear{Chung, Welsh, Barres, and Stevens}{Chung
  et~al\mbox{.}}{2015}]%
        {chung2015glia}
\bibfield{author}{\bibinfo{person}{Won-Suk Chung}, \bibinfo{person}{Christina~A
  Welsh}, \bibinfo{person}{Ben~A Barres}, {and} \bibinfo{person}{Beth
  Stevens}.} \bibinfo{year}{2015}\natexlab{}.
\newblock \showarticletitle{Do glia drive synaptic and cognitive impairment in
  disease?}
\newblock \bibinfo{journal}{\emph{Nature neuroscience}} \bibinfo{volume}{18},
  \bibinfo{number}{11} (\bibinfo{year}{2015}), \bibinfo{pages}{1539--1545}.
\newblock


\bibitem[\protect\citeauthoryear{Clawson, Stewart, Eliasmith, and
  Ferrari}{Clawson et~al\mbox{.}}{[n.d.]}]%
        {RN_39}
\bibfield{author}{\bibinfo{person}{T.~S. Clawson}, \bibinfo{person}{T.~C.
  Stewart}, \bibinfo{person}{C. Eliasmith}, {and} \bibinfo{person}{S.
  Ferrari}.} \bibinfo{year}{[n.d.]}\natexlab{}.
\newblock \showarticletitle{An adaptive spiking neural controller for flapping
  insect-scale robots}. In \bibinfo{booktitle}{\emph{2017 IEEE Symposium Series
  on Computational Intelligence (SSCI)}}. \bibinfo{pages}{1--7}.
\newblock
\urldef\tempurl%
\url{https://doi.org/10.1109/SSCI.2017.8285173}
\showDOI{\tempurl}


\bibitem[\protect\citeauthoryear{Cossart, Aronov, and Yuste}{Cossart
  et~al\mbox{.}}{2003}]%
        {RN197}
\bibfield{author}{\bibinfo{person}{Rosa Cossart}, \bibinfo{person}{Dmitriy
  Aronov}, {and} \bibinfo{person}{Rafael Yuste}.}
  \bibinfo{year}{2003}\natexlab{}.
\newblock \showarticletitle{Attractor dynamics of network UP states in the
  neocortex}.
\newblock \bibinfo{journal}{\emph{Nature}} \bibinfo{volume}{423},
  \bibinfo{number}{6937} (\bibinfo{year}{2003}), \bibinfo{pages}{283--288}.
\newblock
\showISSN{0028-0836}


\bibitem[\protect\citeauthoryear{Crick}{Crick}{1984}]%
        {RN214}
\bibfield{author}{\bibinfo{person}{F Crick}.} \bibinfo{year}{1984}\natexlab{}.
\newblock \showarticletitle{Memory and molecular turnover}.
\newblock \bibinfo{journal}{\emph{Nature}} \bibinfo{volume}{312},
  \bibinfo{number}{5990} (\bibinfo{year}{1984}), \bibinfo{pages}{101}.
\newblock
\showISSN{0028-0836}


\bibitem[\protect\citeauthoryear{Dallérac and Rouach}{Dallérac and
  Rouach}{2016}]%
        {ZN49}
\bibfield{author}{\bibinfo{person}{Glenn Dallérac} {and}
  \bibinfo{person}{Nathalie Rouach}.} \bibinfo{year}{2016}\natexlab{}.
\newblock \showarticletitle{Astrocytes as new targets to improve cognitive
  functions}.
\newblock \bibinfo{journal}{\emph{Progress in Neurobiology}}
  \bibinfo{volume}{144} (\bibinfo{year}{2016}), \bibinfo{pages}{48--67}.
\newblock
\showISSN{0301-0082}
\urldef\tempurl%
\url{http://www.sciencedirect.com/science/article/pii/S0301008215300769}
\showURL{%
\tempurl}


\bibitem[\protect\citeauthoryear{Davies, Srinivasa, Lin, Chinya, Cao, Choday,
  Dimou, Joshi, Imam, and Jain}{Davies et~al\mbox{.}}{2018}]%
        {RN_17}
\bibfield{author}{\bibinfo{person}{Mike Davies}, \bibinfo{person}{Narayan
  Srinivasa}, \bibinfo{person}{Tsung-Han Lin}, \bibinfo{person}{Gautham
  Chinya}, \bibinfo{person}{Yongqiang Cao}, \bibinfo{person}{Sri~Harsha
  Choday}, \bibinfo{person}{Georgios Dimou}, \bibinfo{person}{Prasad Joshi},
  \bibinfo{person}{Nabil Imam}, {and} \bibinfo{person}{Shweta Jain}.}
  \bibinfo{year}{2018}\natexlab{}.
\newblock \showarticletitle{Loihi: A neuromorphic manycore processor with
  on-chip learning}.
\newblock \bibinfo{journal}{\emph{IEEE Micro}} \bibinfo{volume}{38},
  \bibinfo{number}{1} (\bibinfo{year}{2018}), \bibinfo{pages}{82--99}.
\newblock
\showISSN{0272-1732}


\bibitem[\protect\citeauthoryear{De~Pitt{\`a}, Brunel, and
  Volterra}{De~Pitt{\`a} et~al\mbox{.}}{2016}]%
        {de2016astrocytes}
\bibfield{author}{\bibinfo{person}{Maurizio De~Pitt{\`a}},
  \bibinfo{person}{Nicolas Brunel}, {and} \bibinfo{person}{Andrea Volterra}.}
  \bibinfo{year}{2016}\natexlab{}.
\newblock \showarticletitle{Astrocytes: Orchestrating synaptic plasticity?}
\newblock \bibinfo{journal}{\emph{Neuroscience}}  \bibinfo{volume}{323}
  (\bibinfo{year}{2016}), \bibinfo{pages}{43--61}.
\newblock


\bibitem[\protect\citeauthoryear{DeWolf, Stewart~Terrence, Slotine, and
  Eliasmith}{DeWolf et~al\mbox{.}}{2016}]%
        {RN_38}
\bibfield{author}{\bibinfo{person}{Travis DeWolf}, \bibinfo{person}{C.
  Stewart~Terrence}, \bibinfo{person}{Jean-Jacques Slotine}, {and}
  \bibinfo{person}{Chris Eliasmith}.} \bibinfo{year}{2016}\natexlab{}.
\newblock \showarticletitle{A spiking neural model of adaptive arm control}.
\newblock \bibinfo{journal}{\emph{Proceedings of the Royal Society B:
  Biological Sciences}} \bibinfo{volume}{283}, \bibinfo{number}{1843}
  (\bibinfo{year}{2016}), \bibinfo{pages}{20162134}.
\newblock
\urldef\tempurl%
\url{https://doi.org/10.1098/rspb.2016.2134}
\showDOI{\tempurl}


\bibitem[\protect\citeauthoryear{Diehl, Neil, Binas, Cook, Liu, and
  Pfeiffer}{Diehl et~al\mbox{.}}{[n.d.]}]%
        {RN_27}
\bibfield{author}{\bibinfo{person}{Peter~U Diehl}, \bibinfo{person}{Daniel
  Neil}, \bibinfo{person}{Jonathan Binas}, \bibinfo{person}{Matthew Cook},
  \bibinfo{person}{Shih-Chii Liu}, {and} \bibinfo{person}{Michael Pfeiffer}.}
  \bibinfo{year}{[n.d.]}\natexlab{}.
\newblock \showarticletitle{Fast-classifying, high-accuracy spiking deep
  networks through weight and threshold balancing}. In
  \bibinfo{booktitle}{\emph{2015 International Joint Conference on Neural
  Networks (IJCNN)}}. \bibinfo{publisher}{IEEE}, \bibinfo{pages}{1--8}.
\newblock
\showISBNx{1479919608}


\bibitem[\protect\citeauthoryear{Eliasmith}{Eliasmith}{2005}]%
        {RN215}
\bibfield{author}{\bibinfo{person}{Chris Eliasmith}.}
  \bibinfo{year}{2005}\natexlab{}.
\newblock \showarticletitle{A unified approach to building and controlling
  spiking attractor networks}.
\newblock \bibinfo{journal}{\emph{Neural computation}} \bibinfo{volume}{17},
  \bibinfo{number}{6} (\bibinfo{year}{2005}), \bibinfo{pages}{1276--1314}.
\newblock


\bibitem[\protect\citeauthoryear{Fellin, Pascual, Gobbo, Pozzan, Haydon, and
  Carmignoto}{Fellin et~al\mbox{.}}{2004}]%
        {RN200}
\bibfield{author}{\bibinfo{person}{Tommaso Fellin}, \bibinfo{person}{Olivier
  Pascual}, \bibinfo{person}{Sara Gobbo}, \bibinfo{person}{Tullio Pozzan},
  \bibinfo{person}{Philip~G Haydon}, {and} \bibinfo{person}{Giorgio
  Carmignoto}.} \bibinfo{year}{2004}\natexlab{}.
\newblock \showarticletitle{Neuronal synchrony mediated by astrocytic glutamate
  through activation of extrasynaptic NMDA receptors}.
\newblock \bibinfo{journal}{\emph{Neuron}} \bibinfo{volume}{43},
  \bibinfo{number}{5} (\bibinfo{year}{2004}), \bibinfo{pages}{729--743}.
\newblock
\showISSN{0896-6273}


\bibitem[\protect\citeauthoryear{Fields, Araque, Johansen-Berg, Lim, Lynch,
  Nave, Nedergaard, Perez, Sejnowski, and Wake}{Fields et~al\mbox{.}}{2014}]%
        {RN192}
\bibfield{author}{\bibinfo{person}{R~Douglas Fields}, \bibinfo{person}{Alfonso
  Araque}, \bibinfo{person}{Heidi Johansen-Berg}, \bibinfo{person}{Soo-Siang
  Lim}, \bibinfo{person}{Gary Lynch}, \bibinfo{person}{Klaus-Armin Nave},
  \bibinfo{person}{Maiken Nedergaard}, \bibinfo{person}{Ray Perez},
  \bibinfo{person}{Terrence Sejnowski}, {and} \bibinfo{person}{Hiroaki Wake}.}
  \bibinfo{year}{2014}\natexlab{}.
\newblock \showarticletitle{Glial biology in learning and cognition}.
\newblock \bibinfo{journal}{\emph{The neuroscientist}} \bibinfo{volume}{20},
  \bibinfo{number}{5} (\bibinfo{year}{2014}), \bibinfo{pages}{426--431}.
\newblock
\showISSN{1073-8584}


\bibitem[\protect\citeauthoryear{Furber, Galluppi, Temple, and Plana}{Furber
  et~al\mbox{.}}{2014}]%
        {RN_18}
\bibfield{author}{\bibinfo{person}{Steve~B Furber}, \bibinfo{person}{Francesco
  Galluppi}, \bibinfo{person}{Steve Temple}, {and} \bibinfo{person}{Luis~A
  Plana}.} \bibinfo{year}{2014}\natexlab{}.
\newblock \showarticletitle{The spinnaker project}.
\newblock \bibinfo{journal}{\emph{Proc. IEEE}} \bibinfo{volume}{102},
  \bibinfo{number}{5} (\bibinfo{year}{2014}), \bibinfo{pages}{652--665}.
\newblock
\showISSN{0018-9219}


\bibitem[\protect\citeauthoryear{Goldberg, De~Pittà, Volman, Berry, and
  Ben-Jacob}{Goldberg et~al\mbox{.}}{2010}]%
        {RN25}
\bibfield{author}{\bibinfo{person}{Mati Goldberg}, \bibinfo{person}{Maurizio
  De~Pittà}, \bibinfo{person}{Vladislav Volman}, \bibinfo{person}{Hugues
  Berry}, {and} \bibinfo{person}{Eshel Ben-Jacob}.}
  \bibinfo{year}{2010}\natexlab{}.
\newblock \showarticletitle{Nonlinear gap junctions enable long-distance
  propagation of pulsating calcium waves in astrocyte networks}.
\newblock \bibinfo{journal}{\emph{PLoS Comput Biol}} \bibinfo{volume}{6},
  \bibinfo{number}{8} (\bibinfo{year}{2010}), \bibinfo{pages}{e1000909}.
\newblock
\showISSN{1553-7358}


\bibitem[\protect\citeauthoryear{Gutierrez-Galan, Dominguez-Morales,
  Perez-Peña, and Linares-Barranco}{Gutierrez-Galan et~al\mbox{.}}{2019}]%
        {RN_40}
\bibfield{author}{\bibinfo{person}{Daniel Gutierrez-Galan},
  \bibinfo{person}{Juan~Pedro Dominguez-Morales}, \bibinfo{person}{Fernando
  Perez-Peña}, {and} \bibinfo{person}{Alejandro Linares-Barranco}.}
  \bibinfo{year}{2019}\natexlab{}.
\newblock \bibinfo{booktitle}{\emph{NeuroPod: a real-time neuromorphic spiking
  CPG applied to robotics}}.
\newblock


\bibitem[\protect\citeauthoryear{Halassa, Fellin, Takano, Dong, and
  Haydon}{Halassa et~al\mbox{.}}{2007}]%
        {halassa2007synaptic}
\bibfield{author}{\bibinfo{person}{Michael~M Halassa}, \bibinfo{person}{Tommaso
  Fellin}, \bibinfo{person}{Hajime Takano}, \bibinfo{person}{Jing-Hui Dong},
  {and} \bibinfo{person}{Philip~G Haydon}.} \bibinfo{year}{2007}\natexlab{}.
\newblock \showarticletitle{Synaptic islands defined by the territory of a
  single astrocyte}.
\newblock \bibinfo{journal}{\emph{Journal of Neuroscience}}
  \bibinfo{volume}{27}, \bibinfo{number}{24} (\bibinfo{year}{2007}),
  \bibinfo{pages}{6473--6477}.
\newblock


\bibitem[\protect\citeauthoryear{Halassa and Haydon}{Halassa and
  Haydon}{2010}]%
        {ZN55}
\bibfield{author}{\bibinfo{person}{Michael~M. Halassa} {and}
  \bibinfo{person}{Philip~G. Haydon}.} \bibinfo{year}{2010}\natexlab{}.
\newblock \showarticletitle{Integrated Brain Circuits: Astrocytic Networks
  Modulate Neuronal Activity and Behavior}.
\newblock \bibinfo{journal}{\emph{Annual Review of Physiology}}
  \bibinfo{volume}{72}, \bibinfo{number}{1} (\bibinfo{year}{2010}),
  \bibinfo{pages}{335--355}.
\newblock
\urldef\tempurl%
\url{https://doi.org/10.1146/annurev-physiol-021909-135843}
\showDOI{\tempurl}


\bibitem[\protect\citeauthoryear{Han, Chen, Wang, Windrem, Wang, Shanz, Xu,
  Oberheim, Bekar, and Betstadt}{Han et~al\mbox{.}}{2013a}]%
        {ZN45}
\bibfield{author}{\bibinfo{person}{Xiaoning Han}, \bibinfo{person}{Michael
  Chen}, \bibinfo{person}{Fushun Wang}, \bibinfo{person}{Martha Windrem},
  \bibinfo{person}{Su Wang}, \bibinfo{person}{Steven Shanz},
  \bibinfo{person}{Qiwu Xu}, \bibinfo{person}{Nancy~Ann Oberheim},
  \bibinfo{person}{Lane Bekar}, {and} \bibinfo{person}{Sarah Betstadt}.}
  \bibinfo{year}{2013}\natexlab{a}.
\newblock \showarticletitle{Forebrain engraftment by human glial progenitor
  cells enhances synaptic plasticity and learning in adult mice}.
\newblock \bibinfo{journal}{\emph{Cell stem cell}} \bibinfo{volume}{12},
  \bibinfo{number}{3} (\bibinfo{year}{2013}), \bibinfo{pages}{342--353}.
\newblock
\showISSN{1934-5909}


\bibitem[\protect\citeauthoryear{Han, Chen, Wang, Windrem, Wang, Shanz, Xu,
  Oberheim, Bekar, Betstadt, et~al\mbox{.}}{Han et~al\mbox{.}}{2013b}]%
        {han2013}
\bibfield{author}{\bibinfo{person}{Xiaoning Han}, \bibinfo{person}{Michael
  Chen}, \bibinfo{person}{Fushun Wang}, \bibinfo{person}{Martha Windrem},
  \bibinfo{person}{Su Wang}, \bibinfo{person}{Steven Shanz},
  \bibinfo{person}{Qiwu Xu}, \bibinfo{person}{Nancy~Ann Oberheim},
  \bibinfo{person}{Lane Bekar}, \bibinfo{person}{Sarah Betstadt},
  {et~al\mbox{.}}} \bibinfo{year}{2013}\natexlab{b}.
\newblock \showarticletitle{Forebrain engraftment by human glial progenitor
  cells enhances synaptic plasticity and learning in adult mice}.
\newblock \bibinfo{journal}{\emph{Cell stem cell}} \bibinfo{volume}{12},
  \bibinfo{number}{3} (\bibinfo{year}{2013}), \bibinfo{pages}{342--353}.
\newblock


\bibitem[\protect\citeauthoryear{Hebb}{Hebb}{2005}]%
        {RN207}
\bibfield{author}{\bibinfo{person}{Donald~Olding Hebb}.}
  \bibinfo{year}{2005}\natexlab{}.
\newblock \bibinfo{booktitle}{\emph{The organization of behavior: A
  neuropsychological theory}}.
\newblock \bibinfo{publisher}{Psychology Press}.
\newblock
\showISBNx{1135631913}


\bibitem[\protect\citeauthoryear{Hinton, Sejnowski, and Ackley}{Hinton
  et~al\mbox{.}}{1984}]%
        {RN_8}
\bibfield{author}{\bibinfo{person}{Geoffrey~E Hinton},
  \bibinfo{person}{Terrence~J Sejnowski}, {and} \bibinfo{person}{David~H
  Ackley}.} \bibinfo{year}{1984}\natexlab{}.
\newblock \bibinfo{booktitle}{\emph{Boltzmann machines: Constraint satisfaction
  networks that learn}}.
\newblock \bibinfo{publisher}{Carnegie-Mellon University, Department of
  Computer Science Pittsburgh, PA}.
\newblock


\bibitem[\protect\citeauthoryear{Hodgkin and Huxley}{Hodgkin and
  Huxley}{1952}]%
        {RN_24}
\bibfield{author}{\bibinfo{person}{Alan~L Hodgkin} {and}
  \bibinfo{person}{Andrew~F Huxley}.} \bibinfo{year}{1952}\natexlab{}.
\newblock \showarticletitle{A quantitative description of membrane current and
  its application to conduction and excitation in nerve}.
\newblock \bibinfo{journal}{\emph{The Journal of physiology}}
  \bibinfo{volume}{117}, \bibinfo{number}{4} (\bibinfo{year}{1952}),
  \bibinfo{pages}{500}.
\newblock


\bibitem[\protect\citeauthoryear{Hopfield}{Hopfield}{1982}]%
        {RN213}
\bibfield{author}{\bibinfo{person}{John~J Hopfield}.}
  \bibinfo{year}{1982}\natexlab{}.
\newblock \showarticletitle{Neural networks and physical systems with emergent
  collective computational abilities}.
\newblock \bibinfo{journal}{\emph{Proceedings of the national academy of
  sciences}} \bibinfo{volume}{79}, \bibinfo{number}{8} (\bibinfo{year}{1982}),
  \bibinfo{pages}{2554--2558}.
\newblock
\showISSN{0027-8424}


\bibitem[\protect\citeauthoryear{Hopfield}{Hopfield}{1984}]%
        {RN_7}
\bibfield{author}{\bibinfo{person}{John~J Hopfield}.}
  \bibinfo{year}{1984}\natexlab{}.
\newblock \showarticletitle{Neurons with graded response have collective
  computational properties like those of two-state neurons}.
\newblock \bibinfo{journal}{\emph{Proceedings of the national academy of
  sciences}} \bibinfo{volume}{81}, \bibinfo{number}{10} (\bibinfo{year}{1984}),
  \bibinfo{pages}{3088--3092}.
\newblock
\showISSN{0027-8424}


\bibitem[\protect\citeauthoryear{Hwu, Isbell, Oros, and Krichmar}{Hwu
  et~al\mbox{.}}{[n.d.]}]%
        {RN_37}
\bibfield{author}{\bibinfo{person}{T. Hwu}, \bibinfo{person}{J. Isbell},
  \bibinfo{person}{N. Oros}, {and} \bibinfo{person}{J. Krichmar}.}
  \bibinfo{year}{[n.d.]}\natexlab{}.
\newblock \showarticletitle{A self-driving robot using deep convolutional
  neural networks on neuromorphic hardware}. In \bibinfo{booktitle}{\emph{2017
  International Joint Conference on Neural Networks (IJCNN)}}.
  \bibinfo{pages}{635--641}.
\newblock
\showISBNx{2161-4407}
\urldef\tempurl%
\url{https://doi.org/10.1109/IJCNN.2017.7965912}
\showDOI{\tempurl}


\bibitem[\protect\citeauthoryear{Izhikevich}{Izhikevich}{2003}]%
        {RN_25}
\bibfield{author}{\bibinfo{person}{Eugene~M Izhikevich}.}
  \bibinfo{year}{2003}\natexlab{}.
\newblock \showarticletitle{Simple model of spiking neurons}.
\newblock \bibinfo{journal}{\emph{IEEE Transactions on neural networks}}
  \bibinfo{volume}{14}, \bibinfo{number}{6} (\bibinfo{year}{2003}),
  \bibinfo{pages}{1569--1572}.
\newblock
\showISSN{1045-9227}


\bibitem[\protect\citeauthoryear{Jin, Zhang, and Li}{Jin
  et~al\mbox{.}}{[n.d.]}]%
        {RN_29}
\bibfield{author}{\bibinfo{person}{Yingyezhe Jin}, \bibinfo{person}{Wenrui
  Zhang}, {and} \bibinfo{person}{Peng Li}.} \bibinfo{year}{[n.d.]}\natexlab{}.
\newblock \showarticletitle{Hybrid macro/micro level backpropagation for
  training deep spiking neural networks}. In \bibinfo{booktitle}{\emph{Advances
  in Neural Information Processing Systems}}. \bibinfo{pages}{7005--7015}.
\newblock


\bibitem[\protect\citeauthoryear{Kleinfeld and Sompolinsky}{Kleinfeld and
  Sompolinsky}{1988}]%
        {RN205}
\bibfield{author}{\bibinfo{person}{D Kleinfeld} {and} \bibinfo{person}{H
  Sompolinsky}.} \bibinfo{year}{1988}\natexlab{}.
\newblock \showarticletitle{Associative neural network model for the generation
  of temporal patterns. Theory and application to central pattern generators}.
\newblock \bibinfo{journal}{\emph{Biophysical Journal}} \bibinfo{volume}{54},
  \bibinfo{number}{6} (\bibinfo{year}{1988}), \bibinfo{pages}{1039--1051}.
\newblock
\showISSN{0006-3495}


\bibitem[\protect\citeauthoryear{Kovács and Pál}{Kovács and Pál}{2017}]%
        {RN203}
\bibfield{author}{\bibinfo{person}{Adrienn Kovács} {and}
  \bibinfo{person}{Balázs Pál}.} \bibinfo{year}{2017}\natexlab{}.
\newblock \showarticletitle{Astrocyte-dependent slow inward currents (SICs)
  participate in neuromodulatory mechanisms in the pedunculopontine nucleus
  (PPN)}.
\newblock \bibinfo{journal}{\emph{Frontiers in cellular neuroscience}}
  \bibinfo{volume}{11} (\bibinfo{year}{2017}).
\newblock


\bibitem[\protect\citeauthoryear{Lapique}{Lapique}{1907}]%
        {RN_26}
\bibfield{author}{\bibinfo{person}{L Lapique}.}
  \bibinfo{year}{1907}\natexlab{}.
\newblock \showarticletitle{Recherches quantitatives sur l'excitation
  electrique des nerfs traitee comme une polarization}.
\newblock \bibinfo{journal}{\emph{J Physiol Pathol Gen}}  \bibinfo{volume}{9}
  (\bibinfo{year}{1907}), \bibinfo{pages}{620--635}.
\newblock


\bibitem[\protect\citeauthoryear{LeCun, Bengio, and Hinton}{LeCun
  et~al\mbox{.}}{2015}]%
        {RN_4}
\bibfield{author}{\bibinfo{person}{Yann LeCun}, \bibinfo{person}{Yoshua
  Bengio}, {and} \bibinfo{person}{Geoffrey Hinton}.}
  \bibinfo{year}{2015}\natexlab{}.
\newblock \showarticletitle{Deep learning}.
\newblock \bibinfo{journal}{\emph{nature}} \bibinfo{volume}{521},
  \bibinfo{number}{7553} (\bibinfo{year}{2015}), \bibinfo{pages}{436}.
\newblock
\showISSN{1476-4687}


\bibitem[\protect\citeauthoryear{Lee, Panda, Srinivasan, and Roy}{Lee
  et~al\mbox{.}}{2018}]%
        {RN_32}
\bibfield{author}{\bibinfo{person}{Chankyu Lee}, \bibinfo{person}{Priyadarshini
  Panda}, \bibinfo{person}{Gopalakrishnan Srinivasan}, {and}
  \bibinfo{person}{Kaushik Roy}.} \bibinfo{year}{2018}\natexlab{}.
\newblock \showarticletitle{Training deep spiking convolutional neural networks
  with stdp-based unsupervised pre-training followed by supervised
  fine-tuning}.
\newblock \bibinfo{journal}{\emph{Frontiers in neuroscience}}
  \bibinfo{volume}{12} (\bibinfo{year}{2018}).
\newblock


\bibitem[\protect\citeauthoryear{Lind, Brazhe, Jessen, Tan, and Lauritzen}{Lind
  et~al\mbox{.}}{2013}]%
        {RN190}
\bibfield{author}{\bibinfo{person}{Barbara~Lykke Lind},
  \bibinfo{person}{Alexey~R Brazhe}, \bibinfo{person}{Sanne~Barsballe Jessen},
  \bibinfo{person}{Florence~CC Tan}, {and} \bibinfo{person}{Martin~J
  Lauritzen}.} \bibinfo{year}{2013}\natexlab{}.
\newblock \showarticletitle{Rapid stimulus-evoked astrocyte Ca2+ elevations and
  hemodynamic responses in mouse somatosensory cortex in vivo}.
\newblock \bibinfo{journal}{\emph{Proceedings of the National Academy of
  Sciences}} \bibinfo{volume}{110}, \bibinfo{number}{48}
  (\bibinfo{year}{2013}), \bibinfo{pages}{E4678--E4687}.
\newblock
\showISSN{0027-8424}


\bibitem[\protect\citeauthoryear{Maass}{Maass}{1997}]%
        {RN_23}
\bibfield{author}{\bibinfo{person}{Wolfgang Maass}.}
  \bibinfo{year}{1997}\natexlab{}.
\newblock \showarticletitle{Networks of spiking neurons: the third generation
  of neural network models}.
\newblock \bibinfo{journal}{\emph{Neural networks}} \bibinfo{volume}{10},
  \bibinfo{number}{9} (\bibinfo{year}{1997}), \bibinfo{pages}{1659--1671}.
\newblock
\showISSN{0893-6080}


\bibitem[\protect\citeauthoryear{McCulloch and Pitts}{McCulloch and
  Pitts}{1943}]%
        {RN_5}
\bibfield{author}{\bibinfo{person}{Warren~S McCulloch} {and}
  \bibinfo{person}{Walter Pitts}.} \bibinfo{year}{1943}\natexlab{}.
\newblock \showarticletitle{A logical calculus of the ideas immanent in nervous
  activity}.
\newblock \bibinfo{journal}{\emph{The bulletin of mathematical biophysics}}
  \bibinfo{volume}{5}, \bibinfo{number}{4} (\bibinfo{year}{1943}),
  \bibinfo{pages}{115--133}.
\newblock
\showISSN{0007-4985}


\bibitem[\protect\citeauthoryear{McNaughton, Battaglia, Jensen, Moser, and
  Moser}{McNaughton et~al\mbox{.}}{2006}]%
        {RN195}
\bibfield{author}{\bibinfo{person}{Bruce~L McNaughton},
  \bibinfo{person}{Francesco~P Battaglia}, \bibinfo{person}{Ole Jensen},
  \bibinfo{person}{Edvard~I Moser}, {and} \bibinfo{person}{May-Britt Moser}.}
  \bibinfo{year}{2006}\natexlab{}.
\newblock \showarticletitle{Path integration and the neural basis of
  the'cognitive map'}.
\newblock \bibinfo{journal}{\emph{Nature Reviews Neuroscience}}
  \bibinfo{volume}{7}, \bibinfo{number}{8} (\bibinfo{year}{2006}),
  \bibinfo{pages}{663--678}.
\newblock
\showISSN{1471-003X}


\bibitem[\protect\citeauthoryear{Mead}{Mead}{1989}]%
        {RN_3}
\bibfield{author}{\bibinfo{person}{Carver Mead}.}
  \bibinfo{year}{1989}\natexlab{}.
\newblock \bibinfo{booktitle}{\emph{Analog VLSI and neural systems}}.
\newblock \bibinfo{publisher}{Addison-Wesley Longman Publishing Co., Inc.} 371
  pages.
\newblock
\showISBNx{0-201-05992-4}


\bibitem[\protect\citeauthoryear{Menon, Fok, Neckar, Khatib, and Boahen}{Menon
  et~al\mbox{.}}{[n.d.]}]%
        {RN_41}
\bibfield{author}{\bibinfo{person}{Samir Menon}, \bibinfo{person}{Sam Fok},
  \bibinfo{person}{Alex Neckar}, \bibinfo{person}{Oussama Khatib}, {and}
  \bibinfo{person}{Kwabena Boahen}.} \bibinfo{year}{[n.d.]}\natexlab{}.
\newblock \showarticletitle{Controlling articulated robots in task-space with
  spiking silicon neurons}. In \bibinfo{booktitle}{\emph{5th IEEE RAS/EMBS
  International Conference on Biomedical Robotics and Biomechatronics}}.
  \bibinfo{publisher}{IEEE}, \bibinfo{pages}{181--186}.
\newblock
\showISBNx{1479931284}


\bibitem[\protect\citeauthoryear{Merolla, Arthur, Alvarez-Icaza, Cassidy,
  Sawada, Akopyan, Jackson, Imam, Guo, and Nakamura}{Merolla
  et~al\mbox{.}}{2014}]%
        {RN_19}
\bibfield{author}{\bibinfo{person}{Paul~A Merolla}, \bibinfo{person}{John~V
  Arthur}, \bibinfo{person}{Rodrigo Alvarez-Icaza}, \bibinfo{person}{Andrew~S
  Cassidy}, \bibinfo{person}{Jun Sawada}, \bibinfo{person}{Filipp Akopyan},
  \bibinfo{person}{Bryan~L Jackson}, \bibinfo{person}{Nabil Imam},
  \bibinfo{person}{Chen Guo}, {and} \bibinfo{person}{Yutaka Nakamura}.}
  \bibinfo{year}{2014}\natexlab{}.
\newblock \showarticletitle{A million spiking-neuron integrated circuit with a
  scalable communication network and interface}.
\newblock \bibinfo{journal}{\emph{Science}} \bibinfo{volume}{345},
  \bibinfo{number}{6197} (\bibinfo{year}{2014}), \bibinfo{pages}{668--673}.
\newblock
\showISSN{0036-8075}


\bibitem[\protect\citeauthoryear{Navarrete and Araque}{Navarrete and
  Araque}{2010}]%
        {ZN52}
\bibfield{author}{\bibinfo{person}{Marta Navarrete} {and}
  \bibinfo{person}{Alfonso Araque}.} \bibinfo{year}{2010}\natexlab{}.
\newblock \showarticletitle{Endocannabinoids Potentiate Synaptic Transmission
  through Stimulation of Astrocytes}.
\newblock \bibinfo{journal}{\emph{Neuron}} \bibinfo{volume}{68},
  \bibinfo{number}{1} (\bibinfo{year}{2010}), \bibinfo{pages}{113--126}.
\newblock
\showISSN{0896-6273}
\urldef\tempurl%
\url{http://www.sciencedirect.com/science/article/pii/S0896627310006860}
\showURL{%
\tempurl}


\bibitem[\protect\citeauthoryear{P{\'a}l}{P{\'a}l}{2015}]%
        {pal2015astrocytic}
\bibfield{author}{\bibinfo{person}{Bal{\'a}zs P{\'a}l}.}
  \bibinfo{year}{2015}\natexlab{}.
\newblock \showarticletitle{Astrocytic actions on extrasynaptic neuronal
  currents}.
\newblock \bibinfo{journal}{\emph{Frontiers in cellular neuroscience}}
  \bibinfo{volume}{9} (\bibinfo{year}{2015}).
\newblock


\bibitem[\protect\citeauthoryear{Papernot, McDaniel, Jha, Fredrikson, Celik,
  and Swami}{Papernot et~al\mbox{.}}{[n.d.]}]%
        {RN_9}
\bibfield{author}{\bibinfo{person}{Nicolas Papernot}, \bibinfo{person}{Patrick
  McDaniel}, \bibinfo{person}{Somesh Jha}, \bibinfo{person}{Matt Fredrikson},
  \bibinfo{person}{Z~Berkay Celik}, {and} \bibinfo{person}{Ananthram Swami}.}
  \bibinfo{year}{[n.d.]}\natexlab{}.
\newblock \showarticletitle{The limitations of deep learning in adversarial
  settings}. In \bibinfo{booktitle}{\emph{2016 IEEE European Symposium on
  Security and Privacy (EuroS\&P)}}. \bibinfo{publisher}{IEEE},
  \bibinfo{pages}{372--387}.
\newblock
\showISBNx{1509017526}


\bibitem[\protect\citeauthoryear{Parpura and Haydon}{Parpura and
  Haydon}{2000}]%
        {RN113}
\bibfield{author}{\bibinfo{person}{Vladimir Parpura} {and}
  \bibinfo{person}{Philip~G Haydon}.} \bibinfo{year}{2000}\natexlab{}.
\newblock \showarticletitle{Physiological astrocytic calcium levels stimulate
  glutamate release to modulate adjacent neurons}.
\newblock \bibinfo{journal}{\emph{Proceedings of the National Academy of
  Sciences}} \bibinfo{volume}{97}, \bibinfo{number}{15} (\bibinfo{year}{2000}),
  \bibinfo{pages}{8629--8634}.
\newblock
\showISSN{0027-8424}


\bibitem[\protect\citeauthoryear{Perea and Araque}{Perea and Araque}{2010}]%
        {ZN46}
\bibfield{author}{\bibinfo{person}{Gertrudis Perea} {and}
  \bibinfo{person}{Alfonso Araque}.} \bibinfo{year}{2010}\natexlab{}.
\newblock \showarticletitle{GLIA modulates synaptic transmission}.
\newblock \bibinfo{journal}{\emph{Brain research reviews}}
  \bibinfo{volume}{63}, \bibinfo{number}{1} (\bibinfo{year}{2010}),
  \bibinfo{pages}{93--102}.
\newblock
\showISSN{0165-0173}


\bibitem[\protect\citeauthoryear{Perea, Sur, and Araque}{Perea
  et~al\mbox{.}}{2014}]%
        {ZN48}
\bibfield{author}{\bibinfo{person}{Gertrudis Perea}, \bibinfo{person}{Mriganka
  Sur}, {and} \bibinfo{person}{Alfonso Araque}.}
  \bibinfo{year}{2014}\natexlab{}.
\newblock \showarticletitle{Neuron-glia networks: integral gear of brain
  function}.
\newblock \bibinfo{journal}{\emph{Frontiers in Cellular Neuroscience}}
  \bibinfo{volume}{8}, \bibinfo{number}{378} (\bibinfo{year}{2014}).
\newblock
\showISSN{1662-5102}
\urldef\tempurl%
\url{https://doi.org/10.3389/fncel.2014.00378}
\showDOI{\tempurl}


\bibitem[\protect\citeauthoryear{Polykretis, Ivanov, and Michmizos}{Polykretis
  et~al\mbox{.}}{2018a}]%
        {polykretis2018astrocytic}
\bibfield{author}{\bibinfo{person}{Ioannis Polykretis},
  \bibinfo{person}{Vladimir Ivanov}, {and} \bibinfo{person}{Konstantinos~P
  Michmizos}.} \bibinfo{year}{2018}\natexlab{a}.
\newblock \showarticletitle{The astrocytic microdomain as a generative
  mechanism for local plasticity}. In \bibinfo{booktitle}{\emph{International
  Conference on Brain Informatics}}. Springer, \bibinfo{pages}{153--162}.
\newblock


\bibitem[\protect\citeauthoryear{Polykretis, Ivanov, and Michmizos}{Polykretis
  et~al\mbox{.}}{2018b}]%
        {polykretis2018}
\bibfield{author}{\bibinfo{person}{Ioannis Polykretis},
  \bibinfo{person}{Vladimir Ivanov}, {and} \bibinfo{person}{Konstantinos~P
  Michmizos}.} \bibinfo{year}{2018}\natexlab{b}.
\newblock \showarticletitle{A Neural-Astrocytic Network Architecture:
  Astrocytic calcium waves modulate synchronous neuronal activity}. In
  \bibinfo{booktitle}{\emph{Proceedings of the International Conference on
  Neuromorphic Systems}}. \bibinfo{pages}{1--8}.
\newblock


\bibitem[\protect\citeauthoryear{Polykretis, Ivanov, and Michmizos}{Polykretis
  et~al\mbox{.}}{2019}]%
        {polykretis2019}
\bibfield{author}{\bibinfo{person}{Ioannis~E Polykretis},
  \bibinfo{person}{Vladimir~A Ivanov}, {and} \bibinfo{person}{Konstantinos~P
  Michmizos}.} \bibinfo{year}{2019}\natexlab{}.
\newblock \showarticletitle{Computational Astrocyence: Astrocytes encode
  inhibitory activity into the frequency and spatial extent of their calcium
  elevations}. In \bibinfo{booktitle}{\emph{2019 IEEE EMBS International
  Conference on Biomedical \& Health Informatics (BHI)}}. IEEE,
  \bibinfo{pages}{1--4}.
\newblock


\bibitem[\protect\citeauthoryear{Rosenblatt}{Rosenblatt}{1958}]%
        {RN_6}
\bibfield{author}{\bibinfo{person}{Frank Rosenblatt}.}
  \bibinfo{year}{1958}\natexlab{}.
\newblock \showarticletitle{The perceptron: a probabilistic model for
  information storage and organization in the brain}.
\newblock \bibinfo{journal}{\emph{Psychological review}} \bibinfo{volume}{65},
  \bibinfo{number}{6} (\bibinfo{year}{1958}), \bibinfo{pages}{386}.
\newblock
\showISSN{1939-1471}


\bibitem[\protect\citeauthoryear{Rosenfeld, Zemel, and Tsotsos}{Rosenfeld
  et~al\mbox{.}}{2018}]%
        {RN_16}
\bibfield{author}{\bibinfo{person}{Amir Rosenfeld}, \bibinfo{person}{Richard
  Zemel}, {and} \bibinfo{person}{John~K Tsotsos}.}
  \bibinfo{year}{2018}\natexlab{}.
\newblock \showarticletitle{The elephant in the room}.
\newblock \bibinfo{journal}{\emph{arXiv preprint arXiv:1808.03305}}
  (\bibinfo{year}{2018}).
\newblock


\bibitem[\protect\citeauthoryear{Schemmel, Briiderle, Griibl, Hock, Meier, and
  Millner}{Schemmel et~al\mbox{.}}{[n.d.]}]%
        {RN_20}
\bibfield{author}{\bibinfo{person}{Johannes Schemmel}, \bibinfo{person}{Daniel
  Briiderle}, \bibinfo{person}{Andreas Griibl}, \bibinfo{person}{Matthias
  Hock}, \bibinfo{person}{Karlheinz Meier}, {and} \bibinfo{person}{Sebastian
  Millner}.} \bibinfo{year}{[n.d.]}\natexlab{}.
\newblock \showarticletitle{A wafer-scale neuromorphic hardware system for
  large-scale neural modeling}. In \bibinfo{booktitle}{\emph{Proceedings of
  2010 IEEE International Symposium on Circuits and Systems}}.
  \bibinfo{publisher}{IEEE}, \bibinfo{pages}{1947--1950}.
\newblock
\showISBNx{1424453089}


\bibitem[\protect\citeauthoryear{Sejnowski, Churchland, and Movshon}{Sejnowski
  et~al\mbox{.}}{2014}]%
        {RN_43}
\bibfield{author}{\bibinfo{person}{Terrence~J Sejnowski},
  \bibinfo{person}{Patricia~S Churchland}, {and} \bibinfo{person}{J~Anthony
  Movshon}.} \bibinfo{year}{2014}\natexlab{}.
\newblock \showarticletitle{Putting big data to good use in neuroscience}.
\newblock \bibinfo{journal}{\emph{Nature neuroscience}} \bibinfo{volume}{17},
  \bibinfo{number}{11} (\bibinfo{year}{2014}), \bibinfo{pages}{1440}.
\newblock
\showISSN{1546-1726}


\bibitem[\protect\citeauthoryear{Sengupta, Ye, Wang, Liu, and Roy}{Sengupta
  et~al\mbox{.}}{2019}]%
        {RN_31}
\bibfield{author}{\bibinfo{person}{Abhronil Sengupta}, \bibinfo{person}{Yuting
  Ye}, \bibinfo{person}{Robert Wang}, \bibinfo{person}{Chiao Liu}, {and}
  \bibinfo{person}{Kaushik Roy}.} \bibinfo{year}{2019}\natexlab{}.
\newblock \showarticletitle{Going deeper in spiking neural networks: VGG and
  residual architectures}.
\newblock \bibinfo{journal}{\emph{Frontiers in neuroscience}}
  \bibinfo{volume}{13} (\bibinfo{year}{2019}).
\newblock


\bibitem[\protect\citeauthoryear{Shamir}{Shamir}{2018}]%
        {RN_14}
\bibfield{author}{\bibinfo{person}{Ohad Shamir}.}
  \bibinfo{year}{2018}\natexlab{}.
\newblock \showarticletitle{Distribution-specific hardness of learning neural
  networks}.
\newblock \bibinfo{journal}{\emph{The Journal of Machine Learning Research}}
  \bibinfo{volume}{19}, \bibinfo{number}{1} (\bibinfo{year}{2018}),
  \bibinfo{pages}{1135--1163}.
\newblock
\showISSN{1532-4435}


\bibitem[\protect\citeauthoryear{Sharif, Bhagavatula, Bauer, and Reiter}{Sharif
  et~al\mbox{.}}{[n.d.]}]%
        {RN_11}
\bibfield{author}{\bibinfo{person}{Mahmood Sharif}, \bibinfo{person}{Sruti
  Bhagavatula}, \bibinfo{person}{Lujo Bauer}, {and} \bibinfo{person}{Michael~K
  Reiter}.} \bibinfo{year}{[n.d.]}\natexlab{}.
\newblock \showarticletitle{Accessorize to a crime: Real and stealthy attacks
  on state-of-the-art face recognition}. In
  \bibinfo{booktitle}{\emph{Proceedings of the 2016 ACM SIGSAC Conference on
  Computer and Communications Security}}. \bibinfo{publisher}{ACM},
  \bibinfo{pages}{1528--1540}.
\newblock
\showISBNx{145034139X}


\bibitem[\protect\citeauthoryear{Shrestha and Orchard}{Shrestha and
  Orchard}{[n.d.]}]%
        {RN_28}
\bibfield{author}{\bibinfo{person}{Sumit~Bam Shrestha} {and}
  \bibinfo{person}{Garrick Orchard}.} \bibinfo{year}{[n.d.]}\natexlab{}.
\newblock \showarticletitle{SLAYER: Spike layer error reassignment in time}. In
  \bibinfo{booktitle}{\emph{Advances in Neural Information Processing
  Systems}}. \bibinfo{pages}{1412--1421}.
\newblock


\bibitem[\protect\citeauthoryear{Sompolinsky and Kanter}{Sompolinsky and
  Kanter}{1986}]%
        {RN206}
\bibfield{author}{\bibinfo{person}{Haim Sompolinsky} {and} \bibinfo{person}{I
  Kanter}.} \bibinfo{year}{1986}\natexlab{}.
\newblock \showarticletitle{Temporal association in asymmetric neural
  networks}.
\newblock \bibinfo{journal}{\emph{Physical review letters}}
  \bibinfo{volume}{57}, \bibinfo{number}{22} (\bibinfo{year}{1986}),
  \bibinfo{pages}{2861}.
\newblock


\bibitem[\protect\citeauthoryear{Su, Vargas, and Sakurai}{Su
  et~al\mbox{.}}{2019}]%
        {RN_13}
\bibfield{author}{\bibinfo{person}{Jiawei Su},
  \bibinfo{person}{Danilo~Vasconcellos Vargas}, {and} \bibinfo{person}{Kouichi
  Sakurai}.} \bibinfo{year}{2019}\natexlab{}.
\newblock \showarticletitle{One pixel attack for fooling deep neural networks}.
\newblock \bibinfo{journal}{\emph{IEEE Transactions on Evolutionary
  Computation}} (\bibinfo{year}{2019}).
\newblock
\showISSN{1089-778X}


\bibitem[\protect\citeauthoryear{Tang, Kumar, and Michmizos}{Tang
  et~al\mbox{.}}{2020}]%
        {tang2020reinforcement}
\bibfield{author}{\bibinfo{person}{Guangzhi Tang}, \bibinfo{person}{Neelesh
  Kumar}, {and} \bibinfo{person}{Konstantinos~P Michmizos}.}
  \bibinfo{year}{2020}\natexlab{}.
\newblock \showarticletitle{Reinforcement co-Learning of Deep and Spiking
  Neural Networks for Energy-Efficient Mapless Navigation with Neuromorphic
  Hardware}.
\newblock \bibinfo{journal}{\emph{arXiv preprint arXiv:2003.01157}}
  (\bibinfo{year}{2020}).
\newblock


\bibitem[\protect\citeauthoryear{Tang, Polykretis, Ivanov, Shah, and
  Michmizos}{Tang et~al\mbox{.}}{2019a}]%
        {RN_34}
\bibfield{author}{\bibinfo{person}{Guangzhi Tang}, \bibinfo{person}{Ioannis~E.
  Polykretis}, \bibinfo{person}{Vladimir~A. Ivanov}, \bibinfo{person}{Arpit
  Shah}, {and} \bibinfo{person}{Konstantinos~P. Michmizos}.}
  \bibinfo{year}{2019}\natexlab{a}.
\newblock \showarticletitle{Introducing Astrocytes on a Neuromorphic Processor:
  Synchronization, Local Plasticity and Edge of Chaos}.
\newblock \bibinfo{journal}{\emph{ACM Proceedings of 2019 Neuroinspired
  Computing Elements (NICE 2019)}} \bibinfo{volume}{1}, \bibinfo{number}{1}
  (\bibinfo{year}{2019}), \bibinfo{pages}{1--10}.
\newblock
\urldef\tempurl%
\url{https://doi.org/arXiv:1907.01620}
\showDOI{\tempurl}


\bibitem[\protect\citeauthoryear{Tang, Shah, and Michmizos}{Tang
  et~al\mbox{.}}{2019b}]%
        {RN_42}
\bibfield{author}{\bibinfo{person}{Guangzhi Tang}, \bibinfo{person}{Arpit
  Shah}, {and} \bibinfo{person}{Konstantinos~P Michmizos}.}
  \bibinfo{year}{2019}\natexlab{b}.
\newblock \showarticletitle{Spiking neural network on neuromorphic hardware for
  energy-efficient unidimensional SLAM}.
\newblock \bibinfo{journal}{\emph{2019 IEEE/RSJ International Conference on
  Intelligent Robots and Systems (IROS 2019)}} (\bibinfo{year}{2019}),
  \bibinfo{pages}{1--6}.
\newblock
\urldef\tempurl%
\url{https://doi.org/arXiv:1903.02504}
\showDOI{\tempurl}


\bibitem[\protect\citeauthoryear{Tavanaei, Ghodrati, Kheradpisheh, Masquelier,
  and Maida}{Tavanaei et~al\mbox{.}}{2018}]%
        {RN_30}
\bibfield{author}{\bibinfo{person}{Amirhossein Tavanaei},
  \bibinfo{person}{Masoud Ghodrati}, \bibinfo{person}{Saeed~Reza Kheradpisheh},
  \bibinfo{person}{Timothée Masquelier}, {and} \bibinfo{person}{Anthony
  Maida}.} \bibinfo{year}{2018}\natexlab{}.
\newblock \showarticletitle{Deep learning in spiking neural networks}.
\newblock \bibinfo{journal}{\emph{Neural Networks}} (\bibinfo{year}{2018}).
\newblock
\showISSN{0893-6080}


\bibitem[\protect\citeauthoryear{Tewari and Parpura}{Tewari and
  Parpura}{2013}]%
        {ZN51}
\bibfield{author}{\bibinfo{person}{Shivendra Tewari} {and}
  \bibinfo{person}{Vladimir Parpura}.} \bibinfo{year}{2013}\natexlab{}.
\newblock \showarticletitle{A possible role of astrocytes in contextual memory
  retrieval: An analysis obtained using a quantitative framework}.
\newblock \bibinfo{journal}{\emph{Frontiers in Computational Neuroscience}}
  \bibinfo{volume}{7}, \bibinfo{number}{145} (\bibinfo{year}{2013}).
\newblock
\showISSN{1662-5188}
\urldef\tempurl%
\url{https://doi.org/10.3389/fncom.2013.00145}
\showDOI{\tempurl}


\bibitem[\protect\citeauthoryear{Theodosis, Poulain, and Oliet}{Theodosis
  et~al\mbox{.}}{2008a}]%
        {RN216}
\bibfield{author}{\bibinfo{person}{Dionysia~T Theodosis},
  \bibinfo{person}{Dominique~A Poulain}, {and} \bibinfo{person}{Stéphane~HR
  Oliet}.} \bibinfo{year}{2008}\natexlab{a}.
\newblock \showarticletitle{Activity-dependent structural and functional
  plasticity of astrocyte-neuron interactions}.
\newblock \bibinfo{journal}{\emph{Physiological reviews}} \bibinfo{volume}{88},
  \bibinfo{number}{3} (\bibinfo{year}{2008}), \bibinfo{pages}{983--1008}.
\newblock
\showISSN{0031-9333}


\bibitem[\protect\citeauthoryear{Theodosis, Poulain, and Oliet}{Theodosis
  et~al\mbox{.}}{2008b}]%
        {ZN53}
\bibfield{author}{\bibinfo{person}{Dionysia~T. Theodosis},
  \bibinfo{person}{Dominique~A. Poulain}, {and} \bibinfo{person}{Stéphane
  H.~R. Oliet}.} \bibinfo{year}{2008}\natexlab{b}.
\newblock \showarticletitle{Activity-Dependent Structural and Functional
  Plasticity of Astrocyte-Neuron Interactions}.
\newblock \bibinfo{journal}{\emph{Physiological Reviews}} \bibinfo{volume}{88},
  \bibinfo{number}{3} (\bibinfo{year}{2008}), \bibinfo{pages}{983--1008}.
\newblock
\urldef\tempurl%
\url{https://doi.org/10.1152/physrev.00036.2007}
\showDOI{\tempurl}


\bibitem[\protect\citeauthoryear{Turing}{Turing}{1948}]%
        {RN_1}
\bibfield{author}{\bibinfo{person}{Alan~M Turing}.}
  \bibinfo{year}{1948}\natexlab{}.
\newblock \showarticletitle{Intelligent machinery, a heretical theory}.
\newblock \bibinfo{journal}{\emph{The Turing Test: Verbal Behavior as the
  Hallmark of Intelligence}}  \bibinfo{volume}{105} (\bibinfo{year}{1948}).
\newblock


\bibitem[\protect\citeauthoryear{Verkhratsky, Olabarria, Noristani, Yeh, and
  Rodriguez}{Verkhratsky et~al\mbox{.}}{2010}]%
        {verkhratsky2010astrocytes}
\bibfield{author}{\bibinfo{person}{Alexei Verkhratsky}, \bibinfo{person}{Markel
  Olabarria}, \bibinfo{person}{Harun~N Noristani}, \bibinfo{person}{Chia-Yu
  Yeh}, {and} \bibinfo{person}{Jose~Julio Rodriguez}.}
  \bibinfo{year}{2010}\natexlab{}.
\newblock \showarticletitle{Astrocytes in Alzheimer's disease}.
\newblock \bibinfo{journal}{\emph{Neurotherapeutics}} \bibinfo{volume}{7},
  \bibinfo{number}{4} (\bibinfo{year}{2010}), \bibinfo{pages}{399--412}.
\newblock


\bibitem[\protect\citeauthoryear{Volterra and Meldolesi}{Volterra and
  Meldolesi}{2005}]%
        {ZN47}
\bibfield{author}{\bibinfo{person}{Andrea Volterra} {and}
  \bibinfo{person}{Jacopo Meldolesi}.} \bibinfo{year}{2005}\natexlab{}.
\newblock \showarticletitle{Astrocytes, from brain glue to communication
  elements: the revolution continues}.
\newblock \bibinfo{journal}{\emph{Nature Reviews Neuroscience}}
  \bibinfo{volume}{6}, \bibinfo{number}{8} (\bibinfo{year}{2005}),
  \bibinfo{pages}{626}.
\newblock
\showISSN{1471-0048}


\bibitem[\protect\citeauthoryear{Von~Neumann}{Von~Neumann}{2012}]%
        {RN_2}
\bibfield{author}{\bibinfo{person}{John Von~Neumann}.}
  \bibinfo{year}{2012}\natexlab{}.
\newblock \bibinfo{booktitle}{\emph{The computer and the brain}}.
\newblock \bibinfo{publisher}{Yale University Press}.
\newblock
\showISBNx{0300181116}


\bibitem[\protect\citeauthoryear{Wade, McDaid, Harkin, Crunelli, and
  Kelso}{Wade et~al\mbox{.}}{2011}]%
        {RN112}
\bibfield{author}{\bibinfo{person}{John~J Wade}, \bibinfo{person}{Liam~J
  McDaid}, \bibinfo{person}{Jim Harkin}, \bibinfo{person}{Vincenzo Crunelli},
  {and} \bibinfo{person}{JA~Scott Kelso}.} \bibinfo{year}{2011}\natexlab{}.
\newblock \showarticletitle{Bidirectional coupling between astrocytes and
  neurons mediates learning and dynamic coordination in the brain: a multiple
  modeling approach}.
\newblock \bibinfo{journal}{\emph{PloS one}} \bibinfo{volume}{6},
  \bibinfo{number}{12} (\bibinfo{year}{2011}), \bibinfo{pages}{e29445}.
\newblock
\showISSN{1932-6203}


\bibitem[\protect\citeauthoryear{Wimmer, Nykamp, Constantinidis, and
  Compte}{Wimmer et~al\mbox{.}}{2014}]%
        {RN196}
\bibfield{author}{\bibinfo{person}{Klaus Wimmer}, \bibinfo{person}{Duane~Q
  Nykamp}, \bibinfo{person}{Christos Constantinidis}, {and}
  \bibinfo{person}{Albert Compte}.} \bibinfo{year}{2014}\natexlab{}.
\newblock \showarticletitle{Bump attractor dynamics in prefrontal cortex
  explains behavioral precision in spatial working memory}.
\newblock \bibinfo{journal}{\emph{Nature neuroscience}} \bibinfo{volume}{17},
  \bibinfo{number}{3} (\bibinfo{year}{2014}), \bibinfo{pages}{431--439}.
\newblock
\showISSN{1097-6256}


\bibitem[\protect\citeauthoryear{Zhang, Bengio, Hardt, Recht, and
  Vinyals}{Zhang et~al\mbox{.}}{2016}]%
        {RN_12}
\bibfield{author}{\bibinfo{person}{Chiyuan Zhang}, \bibinfo{person}{Samy
  Bengio}, \bibinfo{person}{Moritz Hardt}, \bibinfo{person}{Benjamin Recht},
  {and} \bibinfo{person}{Oriol Vinyals}.} \bibinfo{year}{2016}\natexlab{}.
\newblock \showarticletitle{Understanding deep learning requires rethinking
  generalization}.
\newblock \bibinfo{journal}{\emph{arXiv preprint arXiv:1611.03530}}
  (\bibinfo{year}{2016}).
\newblock


\bibitem[\protect\citeauthoryear{Zhu, Akrout, Zheng, Pelegris, Jayarajan,
  Phanishayee, Schroeder, and Pekhimenko}{Zhu et~al\mbox{.}}{[n.d.]}]%
        {RN_10}
\bibfield{author}{\bibinfo{person}{Hongyu Zhu}, \bibinfo{person}{Mohamed
  Akrout}, \bibinfo{person}{Bojian Zheng}, \bibinfo{person}{Andrew Pelegris},
  \bibinfo{person}{Anand Jayarajan}, \bibinfo{person}{Amar Phanishayee},
  \bibinfo{person}{Bianca Schroeder}, {and} \bibinfo{person}{Gennady
  Pekhimenko}.} \bibinfo{year}{[n.d.]}\natexlab{}.
\newblock \showarticletitle{Benchmarking and analyzing deep neural network
  training}. In \bibinfo{booktitle}{\emph{2018 IEEE International Symposium on
  Workload Characterization (IISWC)}}. \bibinfo{publisher}{IEEE},
  \bibinfo{pages}{88--100}.
\newblock
\showISBNx{1538667800}


\end{thebibliography}

\end{document}